# Scalar induced gravitational waves in chiral scalar–tensor theory of gravity

**Jia-Xi Feng**[1,a], **Fengge Zhang**[1,2,b], **Xian Gao**[1,c]

[1] School of Physics and Astronomy, Sun Yat-sen University, Zhuhai 519082, China
[2] Henan Academy of Sciences, Zhengzhou 450046, Henan, China

**Abstract** We study the scalar induced gravitational waves (SIGWs) from a chiral scalar–tensor theory of gravity. The parity-violating (PV) Lagrangian contains the Chern–Simons (CS) term and PV scalar–tensor terms, which are built of the quadratic Riemann tensor term and first-order derivatives of a scalar field. We consider SIGWs in two cases, in which the semi-analytic expression to calculate SIGWs can be obtained. Then, we calculate the fractional energy density of SIGWs with a monochromatic power spectrum for the curvature perturbation. We find that the SIGWs in chiral scalar–tensor gravity behave differently from those in GR before and after the peak frequency, which results in a large degree of circular polarization.

## 1 Introduction

The detection of gravitational waves (GWs) from binary systems by LIGO/Virgo [1–11] has opened a new window for exploring the nature of gravity. The GWs generated by some cosmological processes [12–15] in the early universe are also a new and unique doorway to the primordial universe. At the second order in cosmological perturbation theory, scalar perturbations could give rise to GWs known as scalar induced gravitational waves (SIGWs) [16–19]. Recently, SIGWs have attracted much attention [20–36], which provide us a novel tool to probe the early universe and the primordial black holes (PBHs) [37–49]. The generation of large SIGWs requires an enhancement of scalar perturbations on small scales, which can be realized in various inflationary models [50–66]. SIGWs in a general cosmological background have been investigated in [67–69]. The effects of primordial non-Gaussianity on SIGWs are also studied in [70–79]. Furthermore, there is evidence suggesting the detection of SIGWs based on the North American Nanohertz Observatory for Gravitational Waves (NANOGrav) 12.5-year data [80–83]. The most recent stochastic GW signal detected by NANOGrav [84,85], as well as the Parkes Pulsar Timing Array (PPTA) [86], European Pulsar Timing Array (EPTA) [87,88], and Chinese Pulsar Timing Array (CPTA) [89] collaborations, could potentially be explained by SIGWs [90–97]. Furthermore, SIGWs may be detected by future GW detectors, including LISA [98,99], DECIGO [100], Taiji [101] and TianQin [102,103], etc.

Considering gravity theories beyond general relativity (GR), especially the gravity theories with parity-violating (PV) terms have attracted much attention in recent years [104–115]. The most frequently studied example of PV gravity is Chern–Simons (CS) gravity [116]. Linear or primordial gravitational waves in CS gravity exhibit the phenomenon of amplitude birefringence [117–120], which is also present in general PV gravity theories [121–125]. In addition, a general class of chiral scalar–tensor theory of gravity has been proposed in Ref. [105], which extends CS gravity by including PV terms with higher order derivatives of the coupled scalar field. Linear or primordial GWs have been extensively studied in this theory [126–133]. In addition to the phenomenon of amplitude birefringence similar to that in CS gravity, one distinguishable characteristic of higher derivatives of the scalar field is that it causes a difference in the velocities of the left and right-hand polarized modes of GWs, which leads to the phenomenon of velocity birefringence of GWs.

The SIGWs in PV gravity [134–139] have been studied extensively. In this work, we focus on the SIGWs from the chiral scalar–tensor theory of gravity. The PV terms in our model consist of the CS term as well as other PV terms that involve the quadratic term of the Riemann tensor and the first derivative of the coupled scalar field. We will derive the equations of motion (EOMs) and power spectra for SIGWs. To be specific, we will investigate the contributions of these PV terms to the SIGWs during the radiation-dominated era. These PV terms cause notable deviations in the fractional energy density and the degree of circular polarization of SIGWs compared to the predictions made by GR.

[a] e-mail: fengjx57@mail2.sysu.edu.cn
[b] e-mail: zhangfengge@hnas.ac.cn
[c] e-mail: gaoxian@mail.sysu.edu.cn (corresponding author)

The paper is organized as follows. In Sect. 2, we review the chiral scalar–tensor theory of gravity and present the quadratic action for tensor perturbations and the cubic action involving one tensor mode and two scalar modes. In Sect. 3, we derive the EOM for SIGWs in the chiral scalar–tensor theory of gravity for the first time. Section 4 focuses on SIGWs during the radiation-dominated era. Specifically, we analyze Green's function and the PV source term in two cases. In Sect. 5, we calculate the fractional energy density and the degree of circular polarization of SIGWs. Furthermore, we evaluate the contribution of PV terms. Finally, we will draw our conclusions in Sect. 6.

Our work includes two appendices, Appendices A and B, which provide related calculations in detail.

## 2 Chiral scalar–tensor theory of gravity

In this section, we first review the chiral scalar–tensor theory of gravity. Taking into account the PV terms that deviate from GR, we then present the quadratic and cubic actions relevant to the SIGWs.

We consider the chiral scalar–tensor theory of gravity, the action of which takes the form [126]

$$S=\frac{1}{2\kappa^2}\int \mathrm{d}^4x\sqrt{-g}\,(R+\mathcal{L}_{\mathrm{PV}})+\int \mathrm{d}^4x\sqrt{-g}\left(-\frac{1}{2}g^{\mu\nu}\partial_\mu\varphi\partial_\nu\varphi-V(\varphi)\right), \tag{1}$$

where $\kappa^2=8\pi G$ and $\mathcal{L}_{\mathrm{PV}}$ is a PV Lagrangian composed of three terms

$$\mathcal{L}_{\mathrm{PV}}=\mathcal{L}_{\mathrm{CS}}+\mathcal{L}_{\mathrm{PV1}}+\mathcal{L}_{\mathrm{PV2}}. \tag{2}$$

In Eq. (2), $\mathcal{L}_{\mathrm{CS}}$ is the CS term given by [140]

$$\mathcal{L}_{\mathrm{CS}}=\frac{1}{4}\vartheta(\varphi)\,{}^*RR,\qquad {}^*RR=\frac{1}{2}\varepsilon^{\mu\nu\rho\sigma}R_{\rho\sigma\alpha\beta}R^{\alpha\beta}{}_{\mu\nu}, \tag{3}$$

where $\varepsilon^{\rho\sigma\alpha\beta}$ is the Levi–Cività tensor defined by the antisymmetric symbol $\epsilon^{\rho\sigma\alpha\beta}$, with $\varepsilon^{\rho\sigma\alpha\beta}=\epsilon^{\rho\sigma\alpha\beta}/\sqrt{-g}$. $\mathcal{L}_{\mathrm{PV1}}$ contains terms built from the Riemann tensor and first-order derivative of the scalar field, as given by [105].

$$\mathcal{L}_{\mathrm{PV1}}=\sum_{\mathcal{A}=1}^{4}a_{\mathcal{A}}(\varphi)L_{\mathcal{A}}, \tag{4}$$

where

$$\begin{aligned}
L_1&=\varepsilon^{\mu\nu\alpha\beta}R_{\alpha\beta\rho\sigma}R_{\mu\nu}{}^{\rho}{}_{\lambda}\varphi^\sigma\varphi^\lambda,\\
L_2&=\varepsilon^{\mu\nu\alpha\beta}R_{\alpha\beta\rho\sigma}R_{\mu\lambda}{}^{\rho\sigma}\varphi_\nu\varphi^\lambda,\\
L_3&=\varepsilon^{\mu\nu\alpha\beta}R_{\alpha\beta\rho\sigma}R^{\sigma}{}_{\nu}\varphi^\rho\varphi_\mu,\\
L_4&=\varepsilon^{\mu\nu\rho\sigma}R_{\rho\sigma\alpha\beta}R^{\alpha\beta}{}_{\mu\nu}\varphi^\lambda\varphi_\lambda,
\end{aligned}$$

with $\varphi^\mu\equiv\nabla^\mu\varphi$. $\mathcal{L}_{\mathrm{PV2}}$ stands for terms involving second-order derivatives of the scalar field, the explicit expression of which can be found in Ref. [105]. In Eq. (1), we also introduce a canonical kinetic term for the scalar field $\varphi$. The contribution from $\mathcal{L}_{\mathrm{PV2}}$ does not introduce new features in SIGWs and brings us much difficulty in calculating SIGWs. In light of this, we only focus on the contributions from $\mathcal{L}_{\mathrm{CS}}$ and $\mathcal{L}_{\mathrm{PV1}}$ in this paper.

In the Newtonian gauge, the perturbed metric reads

$$\mathrm{d}s^2=g_{\mu\nu}\mathrm{d}x^\mu\mathrm{d}x^\nu=-a^2(1+2\phi)\mathrm{d}\eta^2+a^2\left((1-2\psi)\delta_{ij}+\frac{1}{2}h_{ij}\right)\mathrm{d}x^i\,\mathrm{d}x^j, \tag{5}$$

where we neglect the anisotropic stress [19] and do not consider the vector perturbations.

The PV terms do not contribute to the EOMs of the background and linear scalar perturbation. Thus, the EOMs of the background and linear scalar perturbation are the same as those in GR, which can be seen in Appendix A. Furthermore, we conclude that $\phi=\psi$ from Eq. (A7).

Then we calculate the action of SIGWs, from which we can derive the EOM for the SIGWs. Substituting the above metric (5) into the action (1), we can obtain the action of SIGWs

$$S_{\mathrm{GW}}=S^{(2)}_{hh}+S^{(3)}_{ssh}, \tag{6}$$

where $S^{(2)}_{hh}$ and $S^{(3)}_{ssh}$ are the quadratic action for tensor perturbations and the cubic action that contains one tensor mode and two scalar modes, respectively. The quadratic action contains parity-conserved (PC) and PV parts

$$S^{(2)}_{hh}=\frac{1}{2\kappa^2}\int \mathrm{d}^4x\left(\hat{\mathcal{L}}^{(\mathrm{GR})}_{hh}+\hat{\mathcal{L}}^{(\mathrm{PV})}_{hh}\right), \tag{7}$$

where the PC part is the same as that in GR,

$$\hat{\mathcal{L}}^{(\mathrm{GR})}_{hh}=\frac{a^2}{16}\left[\left(h'_{ij}\right)^2-\left(\partial_k h_{ij}\right)^2\right], \tag{8}$$

and the PV parts are as follows

$$\hat{\mathcal{L}}^{(\mathrm{PV})}_{hh}=\frac{a^2}{16}\left[\frac{c_1(\eta)}{aM_{\mathrm{PV}}}\epsilon^{ijk}h'_{il}\partial_j h'^{\,l}_{k}+\frac{c_2(\eta)}{aM_{\mathrm{PV}}}\epsilon^{ijk}\partial^2 h_{il}\partial_j h^{l}_{k}\right], \tag{9}$$

with a prime denoting the derivative with respect to conformal time $\eta$. The coefficients $c_1$ and $c_2$ are[1]

$$\frac{c_1(\eta)}{M_{\rm PV}} = \frac{\vartheta'}{a} - \frac{1}{a^3}\left(4a_1'\varphi'^2 + 8a_1\varphi'\varphi'' - 16a_1\mathcal{H}\varphi'^2\right) - \frac{1}{a^3}\left(2a_2'\varphi'^2 + 4a_2\varphi'\varphi'' - 4a_2\mathcal{H}\varphi'^2\right) - \frac{1}{a^3}\left(a_3'\varphi'^2 + 2a_3\varphi'\varphi'' - 6a_3\mathcal{H}\varphi'^2\right) - \frac{1}{a^3}\left(8a_4'\varphi'^2 + 16a_4\varphi'\varphi'' - 16a_4\mathcal{H}\varphi'^2\right), \tag{10}$$

$$\frac{c_2(\eta)}{M_{\rm PV}} = \frac{\vartheta'}{a} - \frac{1}{a^3}\left(2a_2'\varphi'^2 + 4a_2\varphi'\varphi'' - 4a_2\mathcal{H}\varphi'^2\right) + \frac{1}{a^3}\left(a_3'\varphi'^2 + 2a_3\varphi'\varphi'' - 2a_3\mathcal{H}\varphi'^2\right) - \frac{1}{a^3}\left(8a_4'\varphi'^2 + 16a_4\varphi'\varphi'' - 16a_4\mathcal{H}\varphi'^2\right), \tag{11}$$

and

$$\frac{c_1(\eta) - c_2(\eta)}{M_{\rm PV}} = -\frac{1}{a^3}\left(4a_1'\varphi'^2 + 8a_1\varphi'\varphi'' - 16a_1\mathcal{H}\varphi'^2\right) - \frac{1}{a^3}\left(2a_3'\varphi'^2 + 4a_3\varphi'\varphi'' - 8a_3\mathcal{H}\varphi'^2\right), \tag{12}$$

with $\mathcal{H} = a'/a$.

The cubic action that represents the interaction between the scalar and tensor perturbations also consists of PC and PV parts,

$$S^{(3)}_{ssh} = \frac{1}{2\kappa^2}\int d^4x \left(\hat{\mathcal{L}}^{\rm (GR)}_{ssh} + \hat{\mathcal{L}}^{\rm (PV)}_{ssh}\right). \tag{13}$$

Once again, the contribution from the PC part is the same as that in GR,

$$\hat{\mathcal{L}}^{\rm (GR)}_{ssh} = \frac{1}{2}a^2\left(2\partial_i\psi\partial_j\psi + \kappa^2\partial_i\delta\varphi\partial_j\delta\varphi\right)h^{ij}. \tag{14}$$

The PV contributions can be split into two parts,

$$\hat{\mathcal{L}}^{\rm (PV)}_{ssh} = \hat{\mathcal{L}}^{\rm (CS)}_{ssh} + \hat{\mathcal{L}}^{\rm (PV1)}_{ssh}, \tag{15}$$

where the first part comes from the CS term in action (1),

$$\hat{\mathcal{L}}^{\rm (CS)}_{ssh} = \frac{1}{2}\epsilon^{kl}{}_i\left(2\vartheta_\varphi'\partial_l\delta\varphi\partial_j\partial_k\psi + 2\vartheta_\varphi\partial_l\delta\varphi'\partial_j\partial_k\psi + 2\vartheta_\varphi\partial_l\delta\varphi\partial_j\partial_k\psi'\right)h^{ij}, \tag{16}$$

and the second part is

$$\begin{aligned}
\hat{\mathcal{L}}^{\rm (PV1)}_{ssh} = \frac{\epsilon^{kl}{}_i}{2}\Bigg\{&\partial_\eta\Bigg[16\left(\frac{a_1\mathcal{H}\varphi'}{a^2}\right)\partial_l\delta\varphi\partial_j\partial_k\psi \\
&-4\left(\frac{a_1\mathcal{H}^2}{a^2}\right)\partial_l\delta\varphi\partial_j\partial_k\delta\varphi + 4\left(\frac{a_1\mathcal{H}'}{a^2}\right)\partial_l\delta\varphi\partial_j\partial_k\delta\varphi \\
&+12\left(\frac{a_1\varphi'^2}{a^2}\right)\partial_l\psi\partial_j\partial_k\psi \\
&-4\left(\frac{a_{1\varphi}\varphi'^2}{a^2}\right)'\partial_l\delta\varphi\partial_j\partial_k\psi\Bigg] - 4\left(\frac{a_1\varphi'}{a^2}\right)''\partial_l\delta\varphi\partial_j\partial_k\psi \\
&-12\left(\frac{a_1\varphi'}{a^2}\right)\partial_l\delta\varphi''\partial_j\partial_k\psi \\
&+4\left(\frac{a_1\varphi'}{a^2}\right)\partial_l\delta\varphi\partial_j\partial_k\psi'' - 16\left(\frac{a_1\varphi'}{a^2}\right)'\partial_l\delta\varphi'\partial_j\partial_k\psi \\
&-8\left(\frac{a_1\varphi'}{a^2}\right)\partial_l\delta\varphi'\partial_j\partial_k\psi' + 4\left(\frac{a_1\varphi'}{a^2}\right)\partial^2(\partial_l\delta\varphi\partial_j\partial_k\psi) \\
&+\partial_\eta\Bigg[-8\left(\frac{a_2\varphi'}{a^2}\right)\partial_l\delta\varphi'\partial_j\partial_k\psi \\
&+8\left(\frac{a_2\varphi'^2}{a^2}\right)\partial_l\psi\partial_j\partial_k\psi - 4\left(\frac{a_{2\varphi}\varphi'^2}{a^2}\right)\partial_l\delta\varphi\partial_j\partial_k\psi\Bigg] \\
&+\partial_\eta\Bigg[8\left(\frac{a_3\mathcal{H}\varphi'}{a^2}\right)\partial_l\delta\varphi\partial_j\partial_k\psi \\
&-2\left(\frac{a_3\mathcal{H}^2}{a^2}\right)\partial_l\delta\varphi\partial_j\partial_k\delta\varphi + 2\left(\frac{a_3\mathcal{H}'}{a^2}\right)\partial_l\delta\varphi\partial_j\partial_k\delta\varphi \\
&+2\left(\frac{a_3\varphi'^2}{a^2}\right)\partial_l\psi\partial_j\partial_k\psi\Bigg] \\
&-2\left(\frac{a_3\varphi'}{a^2}\right)''\partial_l\delta\varphi\partial_j\partial_k\psi - 2\left(\frac{a_3\varphi'}{a^2}\right)\partial_l\delta\varphi''\partial_j\partial_k\psi \\
&+2\left(\frac{a_3\varphi'}{a^2}\right)\partial_l\delta\varphi\partial_j\partial_k\psi'' \\
&-4\left(\frac{a_3\varphi'}{a^2}\right)'\partial_l\delta\varphi'\partial_j\partial_k\psi + 2\left(\frac{a_3\varphi'}{a^2}\right)\partial^2(\partial_l\delta\varphi\partial_j\partial_k\psi) \\
&+\partial_\eta\Bigg[-32\left(\frac{a_4\varphi'}{a^2}\right)\partial_l\delta\varphi'\partial_j\partial_k\psi \\
&+32\left(\frac{a_4\varphi'^2}{a^2}\right)\partial_l\psi\partial_j\partial_k\psi \\
&-16\left(\frac{a_{4\varphi}\varphi'^2}{a^2}\right)\partial_l\delta\varphi\partial_j\partial_k\psi\Bigg]\Bigg\}h^{ij},
\end{aligned} \tag{17}$$

where $a_{\mathcal{A}\varphi} = da_{\mathcal{A}}/d\varphi$, $\delta\varphi$ represents the fluctuation of the scalar field $\varphi$, and we have used the relation $\phi = \psi$.

## 3 The equation of motion for SIGWs

In this section, we derive the EOM of SIGWs from the chiral scalar–tensor theory of gravity. Varying the action (6) with respect to the tensor perturbation $h_{ij}$, we obtain the EOM for SIGWs

$$h''_{ij} + 2\mathcal{H}h'_{ij} - \partial^2 h_{ij} + \left\{\frac{\epsilon^{ilk}}{aM_{\rm PV}}\partial_l\left[c_1 h_{jk}'' + \left(\mathcal{H}c_1 + c_1'\right)h'_{jk} - c_2\partial^2 h_{jk}\right]\right\} = 4\mathcal{T}^{lm}_{ij}\mathcal{S}_{lm}, \tag{18}$$

[1] The form is same as that in the EOM for the liner GWs [130].

where $\mathcal{T}_{ij}^{lm}$ is the projection operator.

We define the Fourier transformation of tensor metric perturbations as

$$h_{ij}(\eta, \boldsymbol{x}) = \sum_{A=R,L} \int \frac{\mathrm{d}^3 k}{(2\pi)^{3/2}} e^{\mathrm{i}\boldsymbol{k}\cdot\boldsymbol{x}} p_{ij}^A h_{\boldsymbol{k}}^A(\eta), \tag{19}$$

where $p_{ij}^A$ $(A = R, L)$ are the polarization tensors defined by

$$p_{ij}^R = \frac{1}{\sqrt{2}}(\mathbf{e}_{ij}^+ + \mathrm{i}\mathbf{e}_{ij}^\times), \quad p_{ij}^L = \frac{1}{\sqrt{2}}(\mathbf{e}_{ij}^+ - \mathrm{i}\mathbf{e}_{ij}^\times), \tag{20}$$

with $\mathbf{e}_{ij}^+ = \mathbf{e}_i\mathbf{e}_j - \bar{\mathbf{e}}_i\bar{\mathbf{e}}_j$, $\mathbf{e}_{ij}^\times = \mathbf{e}_i\bar{\mathbf{e}}_j + \bar{\mathbf{e}}_i\mathbf{e}_j$. The definition of the projection operator is

$$\mathcal{T}^{lm}_{\ \ ij}\mathcal{S}_{lm}(\boldsymbol{x}, \eta) = \sum_{A=R,L} \int \frac{\mathrm{d}^3 \boldsymbol{k}}{(2\pi)^{3/2}} e^{\mathrm{i}\boldsymbol{k}\cdot\boldsymbol{x}} p_{ij}^A p^{Alm} \tilde{\mathcal{S}}_{lm}(\boldsymbol{k}, \eta), \tag{21}$$

where $\tilde{\mathcal{S}}_{ij}$ is the Fourier transformation of the source $\mathcal{S}_{ij}$.

Using this decomposition, the EOM for the SIGWs in Fourier space becomes[2]

$$\tilde{h}_{\boldsymbol{k}}^{A''} + \left[(1+\mu_A)k^2 - \frac{B^{A''}}{B^A}\right]\tilde{h}_{\boldsymbol{k}}^A = \frac{4B^A}{z^A}\mathcal{S}_{\boldsymbol{k}}^A, \tag{22}$$

where

$$\tilde{h}_{\boldsymbol{k}}^A = B^A h_{\boldsymbol{k}}^A, \quad B^A = a\sqrt{z^A}, \tag{23}$$

and

$$z^A \equiv 1 - \frac{k\lambda^A c_1}{aM_{\mathrm{PV}}}, \tag{24}$$

$$\mu_A \equiv \frac{\lambda^A k\,(c_1 - c_2)\,/\,(aM_{\mathrm{PV}})}{1 - \lambda^A k c_1/\,(aM_{\mathrm{PV}})} = \frac{1}{z^A}\frac{\lambda^A k\,(c_1 - c_2)}{aM_{\mathrm{PV}}}. \tag{25}$$

Here $\mu_A$ determines the speed of GWs $c_{\mathrm{T}}^A$ with $\left(c_{\mathrm{T}}^A\right)^2 = 1 + \mu_A$. Similar to CS gravity, we need $z^A > 0$ to avoid ghost instabilities [141,142].

The source term on the right-hand side of Eq. (22) can be split into three parts

$$\mathcal{S}_{\boldsymbol{k}}^A = p^{Alm}\tilde{\mathcal{S}}_{lm}(\boldsymbol{k}) = \mathcal{S}_{\boldsymbol{k}}^{A(\mathrm{GR})} + \mathcal{S}_{\boldsymbol{k}}^{A(\mathrm{CS})} + \mathcal{S}_{\boldsymbol{k}}^{A(\mathrm{PV1})}, \tag{26}$$

where

$$\mathcal{S}_{\boldsymbol{k}}^{A(\mathrm{GR})} = \int \frac{\mathrm{d}^3 \boldsymbol{k}'}{(2\pi)^{3/2}} p^{Aij}\, k_i' k_j' \left[2\psi_{\boldsymbol{k}'}\psi_{\boldsymbol{k}-\boldsymbol{k}'} + \kappa^2 \delta\varphi_{\boldsymbol{k}'}\delta\varphi_{\boldsymbol{k}-\boldsymbol{k}'}\right], \tag{27}$$

$$\begin{aligned}\mathcal{S}_{\boldsymbol{k}}^{A(\mathrm{CS})} = &\int \frac{\mathrm{d}^3 \boldsymbol{k}'}{(2\pi)^{3/2}} p^{Aij}\, k_i' k_j' \left(-\frac{\lambda^A k}{a^2}\right) \\ &\times \Big[ \left(\vartheta_\varphi' \delta\varphi_{\boldsymbol{k}'}\psi_{\boldsymbol{k}-\boldsymbol{k}'} + \vartheta_\varphi \delta\varphi_{\boldsymbol{k}'}'\psi_{\boldsymbol{k}-\boldsymbol{k}'} + \vartheta_\varphi \delta\varphi_{\boldsymbol{k}'}\psi_{\boldsymbol{k}-\boldsymbol{k}'}'\right) \\ &+ \left(\boldsymbol{k}' \leftrightarrow \boldsymbol{k}-\boldsymbol{k}'\right)\Big],\end{aligned} \tag{28}$$

and

$$\begin{aligned}
&\mathcal{S}_{\boldsymbol{k}}^{A(\mathrm{PV1})} \\
&= \int \frac{\mathrm{d}^3 \boldsymbol{k}'}{(2\pi)^{3/2}} p^{Aij}\, k_i' k_j' \left(-\frac{\lambda^A k}{a^2}\right) \\
&\times \Bigg\{ \partial_\eta \Bigg[ 8\left(\frac{a_1\mathcal{H}\varphi'}{a^2}\right)\delta\varphi_{\boldsymbol{k}'}\psi_{\boldsymbol{k}-\boldsymbol{k}'} - 2\left(\frac{a_1\mathcal{H}^2}{a^2}\right)\delta\varphi_{\boldsymbol{k}'}\delta\varphi_{\boldsymbol{k}-\boldsymbol{k}'} \\
&+2\left(\frac{a_1\mathcal{H}'}{a^2}\right)\delta\varphi_{\boldsymbol{k}'}\delta\varphi_{\boldsymbol{k}-\boldsymbol{k}'} + 6\left(\frac{a_1\varphi'^2}{a^2}\right)\psi_{\boldsymbol{k}'}\psi_{\boldsymbol{k}-\boldsymbol{k}'} \\
&-2\left(\frac{a_{1\varphi}\varphi'^2}{a^2}\right)'\delta\varphi_{\boldsymbol{k}'}\psi_{\boldsymbol{k}-\boldsymbol{k}'}\Bigg] \\
&-2\left(\frac{a_1\varphi'}{a^2}\right)''\delta\varphi_{\boldsymbol{k}'}\psi_{\boldsymbol{k}-\boldsymbol{k}'} - 6\left(\frac{a_1\varphi'}{a^2}\right)\delta\varphi_{\boldsymbol{k}'}''\psi_{\boldsymbol{k}-\boldsymbol{k}'} \\
&+2\left(\frac{a_1\varphi'}{a^2}\right)\delta\varphi_{\boldsymbol{k}'}\psi_{\boldsymbol{k}-\boldsymbol{k}'}'' \\
&-8\left(\frac{a_1\varphi'}{a^2}\right)'\delta\varphi_{\boldsymbol{k}'}'\psi_{\boldsymbol{k}-\boldsymbol{k}'} - 4\left(\frac{a_1\varphi'}{a^2}\right)\delta\varphi_{\boldsymbol{k}'}'\psi_{\boldsymbol{k}-\boldsymbol{k}'}' \\
&-2\left(\frac{a_1\varphi'}{a^2}\right)(\delta\varphi_{\boldsymbol{k}'}\psi_{\boldsymbol{k}-\boldsymbol{k}'})W(\boldsymbol{k},\boldsymbol{k}') \\
&+\partial_\eta\Bigg[-4\left(\frac{a_2\varphi'}{a^2}\right)\delta\varphi_{\boldsymbol{k}'}'\psi_{\boldsymbol{k}-\boldsymbol{k}'} + 4\left(\frac{a_2\varphi'^2}{a^2}\right)\psi_{\boldsymbol{k}'}\psi_{\boldsymbol{k}-\boldsymbol{k}'} \\
&-2\left(\frac{a_{2\varphi}\varphi'^2}{a^2}\right)\delta\varphi_{\boldsymbol{k}'}\psi_{\boldsymbol{k}-\boldsymbol{k}'}\Bigg] \\
&+\partial_\eta\Bigg[4\left(\frac{a_3\mathcal{H}\varphi'}{a^2}\right)\delta\varphi_{\boldsymbol{k}'}\psi_{\boldsymbol{k}-\boldsymbol{k}'} - \left(\frac{a_3\mathcal{H}^2}{a^2}\right)\delta\varphi_{\boldsymbol{k}'}\delta\varphi_{\boldsymbol{k}-\boldsymbol{k}'} \\
&+\left(\frac{a_3\mathcal{H}'}{a^2}\right)\delta\varphi_{\boldsymbol{k}'}\delta\varphi_{\boldsymbol{k}-\boldsymbol{k}'} \\
&+\left(\frac{a_3\varphi'^2}{a^2}\right)\psi_{\boldsymbol{k}'}\psi_{\boldsymbol{k}-\boldsymbol{k}'}\Bigg] - \left(\frac{a_3\varphi'}{a^2}\right)''\delta\varphi_{\boldsymbol{k}'}\psi_{\boldsymbol{k}-\boldsymbol{k}'} \\
&-\left(\frac{a_3\varphi'}{a^2}\right)\delta\varphi_{\boldsymbol{k}'}''\psi_{\boldsymbol{k}-\boldsymbol{k}'} \\
&+\left(\frac{a_3\varphi'}{a^2}\right)\delta\varphi_{\boldsymbol{k}'}\psi_{\boldsymbol{k}-\boldsymbol{k}'}'' - 2\left(\frac{a_3\varphi'}{a^2}\right)'\delta\varphi_{\boldsymbol{k}'}'\psi_{\boldsymbol{k}-\boldsymbol{k}'} \\
&-\left(\frac{a_3\varphi'}{a^2}\right)(\delta\varphi_{\boldsymbol{k}'}\psi_{\boldsymbol{k}-\boldsymbol{k}'})W(\boldsymbol{k},\boldsymbol{k}') \\
&+\partial_\eta\Bigg[-16\left(\frac{a_4\varphi'}{a^2}\right)\delta\varphi_{\boldsymbol{k}'}'\psi_{\boldsymbol{k}-\boldsymbol{k}'} + 16\left(\frac{a_4\varphi'^2}{a^2}\right)\psi_{\boldsymbol{k}'}\psi_{\boldsymbol{k}-\boldsymbol{k}'}
\end{aligned}$$

[2] We have used the relation $\epsilon^{ilk}k_l p_{jk}^A = \mathrm{i}k\lambda^A p_j^{i\ A}$ [126] to derive the following EOMs.

$$-8\left(\frac{a_{4\varphi}\varphi'^2}{a^2}\right)\delta\varphi_{\boldsymbol{k}'}\psi_{\boldsymbol{k}-\boldsymbol{k}'}\Big]\Big\}$$
$$+\left(\boldsymbol{k}'\leftrightarrow\boldsymbol{k}-\boldsymbol{k}'\right), \tag{29}$$

with $W(\boldsymbol{k},\boldsymbol{k}')=\boldsymbol{k}'\cdot\boldsymbol{k}'+2\boldsymbol{k}'\cdot(\boldsymbol{k}-\boldsymbol{k}')+(\boldsymbol{k}-\boldsymbol{k}')\cdot(\boldsymbol{k}-\boldsymbol{k}')$, and $\psi_{\boldsymbol{k}}$ and $\delta\varphi_{\boldsymbol{k}}$ are the Fourier modes of the corresponding perturbations.

By the method of the Green's function, the solution of Eq. (22) is given as follows

$$h_{\boldsymbol{k}}^A(\eta)=\frac{4}{B^A(k,\eta)}\int^{\eta}d\bar{\eta}\,G_{\boldsymbol{k}}^A(\eta,\bar{\eta})\frac{B^A(k,\bar{\eta})}{z^A(k,\bar{\eta})}\mathcal{S}_k^{(A)}, \tag{30}$$

where $G_k^A$ is the Green's function, which satisfies

$$G_{\boldsymbol{k}}^{A''}(\eta,\bar{\eta})+\left[(1+\mu_A)k^2-\frac{B^{A''}}{B^A}\right]G_{\boldsymbol{k}}^A(\eta,\bar{\eta})=\delta(\eta-\bar{\eta}). \tag{31}$$

In Eq. (30), the source term is determined by several factors, including the background quantities $\varphi'$, $a(\eta)$, scalar perturbations $\psi_{\boldsymbol{k}}$ and $\delta\varphi_{\boldsymbol{k}}$, as well as coupling functions such as $\vartheta$ and $a_{\mathcal{A}}$. Additionally, Green's function is also affected by $z^A$ and $\mu_A$. To calculate SIGWs, we will further investigate these factors in the next section.

## 4 SIGWs during radiation dominated era

In this section, we will consider the SIGWs in the chiral scalar–tensor theory of gravity during the radiation-dominated era.

### 4.1 The background quantities and first-order perturbations

During the radiation-dominated era, the equation of state is $w=\bar{P}/\bar{\rho}=1/3$. Combing it with the background equations (A1), (A2), we obtain

$$a(\eta)=a_0\eta,\quad \varphi'=\pm\frac{2}{\kappa}\eta^{-1} \tag{32}$$

for the evolution of the scale factor $a$ and $\varphi'$, which yield

$$\mathcal{H}=\frac{1}{\eta},\quad \varphi=\pm\frac{2}{\kappa}\ln(\eta/\eta_0)+\varphi_0, \tag{33}$$

where $\varphi_0$ is the value of $\varphi$ at $\eta_0$.

With Eq. (A5), we can express the fluctuation of the scalar field in terms of the scalar metric perturbations as follows

$$\delta\varphi=\frac{\epsilon^S}{\kappa}\left(\psi+\eta\psi'\right),\quad (\epsilon^S=\pm 1) \tag{34}$$

For later convenience, we split scalar perturbations into the primordial perturbation and the transfer function

$$\psi_{\boldsymbol{k}}=\frac{2}{3}T_\psi(x)\zeta(\boldsymbol{k}), \tag{35}$$

where $x=k\eta$. Note that the PV terms do not alter the evolution of the background and the linear scalar perturbations, thus the transfer function $T_\psi(x)$ is the same as that in GR [23],

$$T_\psi(x)=\frac{9}{x^2}\left(\frac{\sin(x/\sqrt{3})}{x/\sqrt{3}}-\cos(x/\sqrt{3})\right), \tag{36}$$

and the primordial value of $\zeta(\boldsymbol{k})$ is related to the power spectrum of primordial curvature perturbation as

$$\left\langle\zeta(\boldsymbol{k})\,\zeta\left(\boldsymbol{k}'\right)\right\rangle=\delta(\boldsymbol{k}+\boldsymbol{k}')\frac{2\pi}{k^3}\mathcal{P}_\zeta(k). \tag{37}$$

### 4.2 The Green's function

In order to calculate the SIGWs, we should first solve Eq. (31) to obtain the Green's function. $z^A$ and $\mu_A$ in Eq. (31) characterize the deviation from GR and make the Green's function different from that in GR.

Combining Eqs. (10)–(12) and (32), $z^A$ in Eq. (24) and $\mu_A$ in Eq. (25) can be written as

$$z^A=1-k\lambda^A\left[\left(\frac{\vartheta'}{a_0^2\eta^2}\right)-4\left(\frac{2a_2'+8a_4'+4a_1'+a_3'}{\kappa^2a_0^4\eta^6}\right)\right.$$
$$\left.+4\left(\frac{8a_2+32a_4+24a_1+8a_3}{\kappa^2a_0^4\eta^7}\right)\right], \tag{38}$$

and

$$\mu_A=-\frac{k\lambda^A}{z^A}\left[\frac{4(4a_1'+2a_3')}{\kappa^2a_0^4\eta^6}-\frac{4(24a_1+12a_3)}{\kappa^2a_0^4\eta^7}\right]. \tag{39}$$

For general $z^A$ and $\mu_A$, both of which depend on time and wavenumber, it is difficult to obtain the analytic solution for $G_k^A(\eta,\bar{\eta})$. Thus, we make some reasonable assumptions on $\mu_A$ and $z^A$ in light of the observations to simplify the following calculations.

From the EOM for SIGWs (22), $\mu_A$ represents the deviation of the propagating speed of GWs from that of light; thus, it can be constrained by the observation of GWs. The speed of GWs is limited by $-3\times10^{-15}<c_{\mathrm{T}}-1<7\times10^{-16}$ [143], which imposes constraint on $\mu_A$ that

$$-6\times10^{-15}<\mu_A<14\times10^{-16}. \tag{40}$$

Therefore, $\mu_A$ must be very small and negligible. From now on, we assume $\mu_A = 0$. Furthermore, this assumption also constrains the coefficients defined in Eqs. (10) and (11). From Eq. (25), we conclude that

$$\mu_A \propto (c_1 - c_2) = 0. \tag{41}$$

Although we assume $\mu_A = 0$ by taking the observational constraint on the speed of GWs into account, it is still difficult to obtain an analytic expression for the Green's function. In this paper, we mainly concentrate on the contribution from the PV source, so we expect a minimal deviation of the Green's function from that in GR. From the definition of $z^A$ (23), if $z^A$ is also independent of time, then $B^{A''}/B^A = a''/a$, and we can obtain the analytic solution for the Green's function with the condition $\mu_A = 0$. We will see that an exponential form for the coupling functions given by

$$\vartheta = \vartheta_0 \exp^{\kappa\alpha\varphi}, \quad a_{\mathcal{A}} = \mathcal{C}_{\mathcal{A}} \exp^{\kappa\beta_{\mathcal{A}}\varphi}, \quad (\mathcal{A} = 1, 2, 3, 4) \tag{42}$$

can make $z^A$ independent of time and $\mu_A = 0$. In the following we will look for conditions that satisfy the above requirement.

Substituting Eq. (42) into Eqs. (38), (39), we have[3]

$$\begin{aligned}
z^A \;=\; & 1 - k\lambda^A\Bigg[\left(\frac{2\alpha\epsilon^S\vartheta_0\exp^{\kappa\alpha\varphi_0}}{a_0^2\eta_0^{2\alpha\epsilon^S}}\right)\eta^{2\alpha\epsilon^S-3} \\
& - 4(2\beta_2\epsilon^S-4)\left(\frac{2\mathcal{C}_2\exp^{\kappa\beta_2\varphi_0}}{\kappa^2a_0^4\eta_0^{2\beta_2\epsilon^S}}\right)\eta^{2\beta_2\epsilon^S-7} \\
& - 4(2\beta_4\epsilon^S-4)\left(\frac{8\mathcal{C}_4\exp^{\kappa\beta_4\varphi_0}}{\kappa^2a_0^4\eta_0^{2\beta_4\epsilon^S}}\right)\eta^{2\beta_4\epsilon^S-7} \\
& + 4(2\beta_3\epsilon^S-4)\left(\frac{\mathcal{C}_3\exp^{\kappa\beta_3\varphi_0}}{\kappa^2a_0^4\eta_0^{2\beta_3\epsilon^S}}\right)\eta^{2\beta_3\epsilon^S-7}\Bigg], \\
\mu_A = \; & \frac{4k\lambda^A}{z^A}\Bigg[(2\beta_1\epsilon^S-6)\left(\frac{4\mathcal{C}_1\exp^{\kappa\beta_1\varphi_0}}{\kappa^2a_0^4\eta_0^{2\beta_1\epsilon^S}}\right)\eta^{2\beta_1\epsilon^S-7} \\
& + (2\beta_3\epsilon^S-6)\left(\frac{2\mathcal{C}_3\exp^{\kappa\beta_3\varphi_0}}{\kappa^2a_0^4\eta_0^{2\beta_3\epsilon^S}}\right)\eta^{2\beta_3\epsilon^S-7}\Bigg].
\end{aligned} \tag{43}$$

From the above equations, we observe that there are two cases in which $z^A$ is independent of time and $\mu_A = 0$,

- Case 1:

$$2\alpha\epsilon^S - 3 = 0, \quad 2\beta_{\mathcal{A}}\epsilon^S = 7, \quad \text{and} \quad 2\mathcal{C}_1 + \mathcal{C}_3 = 0, \tag{44}$$

- Case 2:

$$2\alpha\epsilon^S - 3 = 0, \quad 2\beta_{\mathcal{A}}\epsilon^S = 4, \quad \text{and} \quad 2\mathcal{C}_1 + \mathcal{C}_3 = 0. \tag{45}$$

In both cases, the Green's function is the same as that in GR,

$$G_{\boldsymbol{k}}^A(\eta, \bar{\eta}) = \frac{\sin\left[k\,(\eta - \bar{\eta})\right]}{k}\Theta(\eta - \bar{\eta}). \tag{46}$$

However, we will see that the source terms are different in these two cases in the next subsection.

### 4.3 The source term

Substituting Eq. (35) into Eq. (27), we obtain the explicit expression for the source term during radiation dominated era,

$$\mathcal{S}_{\boldsymbol{k}}^{A(\mathrm{GR})} = \int \frac{\mathrm{d}^3\boldsymbol{k}'}{(2\pi)^{3/2}}\, p^{Aij}\, k_i'k_j'\zeta(\boldsymbol{k}')\zeta(\boldsymbol{k}-\boldsymbol{k}')\cdot f_{\mathrm{GR}}(u, v, x), \tag{47}$$

with

$$\begin{aligned}
& f_{\mathrm{GR}}(u, v, x) \\
& = \frac{4}{9}\Big(2T_\psi(ux)T_\psi(vx) + [T_\psi(ux) + uxT_\psi^*(ux)] \\
& \quad \times[T_\psi(vx) + vxT_\psi^*(ux)]\Big),
\end{aligned} \tag{48}$$

where $u \equiv k'/k$, $v \equiv |\boldsymbol{k}-\boldsymbol{k}'|/k$, and the $*$ denotes the derivative with respect to the argument. For later convenience, we have also symmetrized the function $f_{\mathrm{GR}}(u, v, x)$ with respect to $u \leftrightarrow v$. Substituting Eqs. (34), (35) into Eqs. (28), (29), we obtain

$$\begin{aligned}
\mathcal{S}_{\boldsymbol{k}}^{A(\mathrm{PV})} = & \int \frac{\mathrm{d}^3\boldsymbol{k}'}{(2\pi)^{3/2}}\, p^{Aij}\, k_i'k_j'\zeta(\boldsymbol{k}')\zeta(\boldsymbol{k}-\boldsymbol{k}') \\
& \times\Big(f_{\mathrm{CS}}^A(k, u, v, x) + f_{\mathrm{PV1}}^A(k, u, v, x)\Big)
\end{aligned}$$

where

$$\begin{aligned}
f_{\mathrm{CS}}^A(k, u, v, x) = & -\frac{4k^2\lambda^A}{9a^2}\cdot\partial_x\Bigg[\vartheta_\varphi\cdot\frac{\epsilon^S}{\kappa}\cdot\Big(T_\psi(ux) \\
& + uxT_\psi^*(ux)\Big)T_\psi(vx)\Bigg] + u \leftrightarrow v.
\end{aligned} \tag{49}$$

As mentioned in the above subsection, there are two cases that can keep $z^A$ independent of time meanwhile $\mu_A = 0$ with the exponential form of the coupling functions (42), so that we can obtain the analytic solution for the Green's function. The function $f^A_{\rm PV1}(k,u,v,x)$ in these two cases has different forms, and we will give the expression of $f^A_{\rm PV1}(k,u,v,x)$ in these two cases, respectively.

[3] Note that, here we have used $\mu_A \propto (c_1 - c_2) = 0$ to derive the expression of $z^A$.

- Case 1: $2\alpha\epsilon^S-3=0$, $2\beta_{\mathcal{A}}\epsilon^S=7$ and $2\mathcal{C}_1+\mathcal{C}_3=0$
  In this case,

$$z^A(k)=1-k\lambda^A\left(\mathcal{D}_{cs}+\mathcal{D}_0\right),\quad \mu_A=0, \tag{50}$$

where

$$\mathcal{D}_{cs}=\frac{3\vartheta_0\exp^{\kappa\alpha\varphi_0}}{a_0^2\eta_0^3},\ \mathcal{D}_0=-\frac{12(2\mathcal{C}_2+8\mathcal{C}_4-\mathcal{C}_3)\exp^{\kappa\beta_2\varphi_0}}{\kappa^2a_0^4\eta_0^7}. \tag{51}$$

In this case, the requirement of $z^A>0$ to avoid ghost field can be satisfied only if $k\left(\mathcal{D}_{cs}+\mathcal{D}_0\right)<1$. Meanwhile, the function $f^A_{\rm PV1}(k,u,v,x)$ has the following form

$$\begin{aligned}
&f^A_{\rm PV1}(k,u,v,x)\\
&=-\frac{4k\lambda^A}{9a^2}\left\{k\cdot\partial_x\left[\left(\frac{2(2a_2+8a_4-a_3)\varphi'^2}{a^2}\right)\right.\right.\\
&\left.\times T_\psi(ux)T_\psi(vx)\right]-k\cdot\partial_x\left[\left(\frac{2(2a_2+8a_4-a_3)\varphi'}{a^2}\right)uk\right.\\
&\left.\times\left[2T^*_\psi(ux)+uxT^{**}_\psi(ux)\right]T_\psi(vx)\right]\\
&-k\cdot\partial_x\left[\left(\frac{(2a_{2\varphi}+8a_{4\varphi}-a_{3\varphi})\varphi'^2}{a^2}\right)\right.\\
&\left.\left.\times\left[T_\psi(ux)+uxT^*_\psi(ux)\right]T_\psi(vx)\right]\right\}+(u\leftrightarrow v)\,.
\end{aligned}$$

With the expressions for the coupling functions in Eq. (42), Eqs. (49) and (52) can be rewritten as

$$\begin{aligned}
&f^A_{\rm CS}(k,u,v,x)\\
&=-\frac{4\mathcal{D}_{cs}k\lambda^A}{9}\left\{3T_\psi(ux)T_\psi(vx)\right.\\
&+3uxT^*_\psi(ux)T_\psi(vx)+3vxT_\psi(ux)T^*_\psi(vx)\\
&+uvx^2T^*_\psi(ux)T^*_\psi(vx)+\frac{1}{2}u^2x^2T^{**}_\psi(ux)T_\psi(vx)\\
&\left.+\frac{1}{2}v^2x^2T_\psi(ux)T^{**}_\psi(vx)\right\},
\end{aligned}\tag{52}$$

and

$$\begin{aligned}
&f^A_{\rm PV1}(k,u,v,x)\\
&=-\frac{4\mathcal{D}_0k\lambda^A}{9}\left\{3T_\psi(ux)T_\psi(vx)+\frac{25}{3}uxT^*_\psi(ux)T_\psi(vx)\right.\\
&+\frac{25}{3}vxT_\psi(ux)T^*_\psi(vx)+\frac{11}{3}uvx^2T^*_\psi(ux)T^*_\psi(vx)\\
&+\frac{7}{2}u^2x^2T^{**}_\psi(ux)T_\psi(vx)+\frac{7}{2}v^2x^2T_\psi(ux)T^{**}_\psi(vx)\\
&+\frac{1}{3}u^2vx^3T^{**}_\psi(ux)T^*_\psi(vx)+\frac{1}{3}uv^2x^3T^*_\psi(ux)T^{**}_\psi(vx)\\
&\left.+\frac{1}{3}u^3x^3T^{***}_\psi(ux)T_\psi(vx)+\frac{1}{3}v^3x^3T_\psi(ux)T^{***}_\psi(vx)\right\}.
\end{aligned}\tag{53}$$

- Case 2: $2\alpha\epsilon^S-3=0$, $2\beta_{\mathcal{A}}\epsilon^S=4$ and $2\mathcal{C}_1+\mathcal{C}_3=0$
  In this case,

$$z^A(k)=1-k\lambda^A\mathcal{D}_{cs},\quad \mu_A=0. \tag{54}$$

Similarly, we require $k\mathcal{D}_{cs}<1$ to avoid the ghost field. In this case, the function $f^A_{\rm PV1}(k,u,v,x)$ is

$$\begin{aligned}
&f^A_{\rm PV1}(k,u,v,x)\\
&=-\frac{4k\lambda^A}{9}\frac{\mathcal{D}_0k^3\eta_0^3}{x^3}\left\{\frac{4}{3}uxT^*_\psi(ux)T_\psi(vx)\right.\\
&+\frac{4}{3}vxT_\psi(ux)T^*_\psi(vx)+\frac{8}{3}uvx^2T^*_\psi(ux)T^*_\psi(vx)\\
&+2u^2x^2T^{**}_\psi(ux)T_\psi(vx)+2v^2x^2T_\psi(ux)T^{**}_\psi(vx)\\
&+\frac{1}{3}u^2vx^3T^{**}_\psi(ux)T^*_\psi(vx)\\
&+\frac{1}{3}u^3x^3T^{***}_\psi(ux)T_\psi(vx)+\frac{1}{3}v^3x^3T_\psi(ux)T^{***}_\psi(vx)\\
&\left.+\frac{1}{3}uv^2x^3T^*_\psi(ux)T^{**}_\psi(vx)\right\},
\end{aligned}\tag{55}$$

and the function $f^A_{\rm CS}(k,u,v,x)$ is the same as the former case.

We observe that when $2\mathcal{C}_2+8\mathcal{C}_4-\mathcal{C}_3=0$ in the above cases, the function $f^A_{\rm PV1}(k,u,v,x)$ vanishes, $z^A(k)$ and the source term take the same form as in CS theory. In this paper, we focus on the PV contributions from both the CS term as well as $\mathcal{L}_{\rm PV1}$ to the power spectra and energy density of the SIGWs; thus, we assume $2\mathcal{C}_2+8\mathcal{C}_4-\mathcal{C}_3\neq0$.

## 5 The power spectra and the degree of the circular polarization

The power spectra $\mathcal{P}^A_h(k,\eta)$ are related to the expectation values as

$$\left\langle h^A_{\boldsymbol{k}}(\eta)h^{A'}_{\boldsymbol{k}'}(\eta)\right\rangle=\frac{2\pi^2}{k^3}\delta^{AA'}\delta^3(\boldsymbol{k}+\boldsymbol{k}')\mathcal{P}^A_h(k,\eta). \tag{56}$$

After some lengthy but straightforward calculations, we obtain the power spectra of SIGWs

$$\mathcal{P}_h^A(k,\eta) = \frac{4}{\left(z^A(k)\right)^2}\int_0^\infty \mathrm{d}u \int_{|1-u|}^{1+u}\mathrm{d}v \times \left[\frac{4u^2-(1+u^2-v^2)^2}{4uv}\right]^2 \times I^A(k,u,v,x)^2\mathcal{P}_\zeta(uk)\mathcal{P}_\zeta(vk), \tag{57}$$

where

$$\begin{aligned} I^A(k,u,v,x) &= \int_0^x \mathrm{d}\bar{x}\; k\,\frac{a(\bar{\eta})}{a(\eta)}G_{\boldsymbol{k}}^A(x,\bar{x})\,\big(f_{\mathrm{GR}}(u,v,\bar{x}) \\ &\quad + f_{\mathrm{CS}}^A(k,u,v,\bar{x}) + f_{\mathrm{PV1}}^A(k,u,v,\bar{x})\big) \\ &= I_{\mathrm{GR}}(u,v,x) + I_{\mathrm{CS}}^A(k,u,v,x) + I_{\mathrm{PV1}}^A(k,u,v,x). \end{aligned} \tag{58}$$

In the above, $I_{\mathrm{GR}}$ is the same as that in GR [23]. The analytic expression of $I^A$ is shown in Appendix B.

The fractional energy density of the SIGWs is

$$\begin{aligned} \Omega_{\mathrm{GW}}(k,x) &= \frac{1}{48}\left(\frac{k}{\mathcal{H}}\right)^2 \sum_{A=R,L}\overline{\mathcal{P}_h^A(k,x)} \\ &= \frac{x^2}{48}\sum_{A=R,L}\overline{\mathcal{P}_h^A(k,x)} \\ &= \frac{1}{12}\int_0^\infty \mathrm{d}u\int_{|1-u|}^{1+u}\mathrm{d}v\,\mathcal{J}(u,v) \\ &\quad \times \sum_{A=R,L}\frac{\overline{\tilde{I}^A(k,u,v,x)^2}}{(z^A(k))^2}\mathcal{P}_\zeta(uk)\mathcal{P}_\zeta(vk), \end{aligned} \tag{59}$$

where the overline represents the time average, and

$$\mathcal{J}(u,v) = \left[\frac{4u^2-(1+u^2-v^2)^2}{4uv}\right]^2, \tag{60}$$

and $\overline{\tilde{I}^A(k,u,v,x)^2} = \overline{I^A(k,u,v,x)^2}x^2$. The GWs behave as free radiation, thus the fractional energy density of the SIGWs at the present time $\Omega_{\mathrm{GW},0}$ can be expressed as [71]

$$\Omega_{\mathrm{GW},0}(k) = \Omega_{\mathrm{GW}}(k,\eta\to\infty)\,\Omega_{r,0}, \tag{61}$$

where $\Omega_{r,0}$ is the current fractional energy density of the radiation.

The degree of the circular polarization is defined as [144, 145]

$$\Pi \equiv \frac{\overline{\mathcal{P}_h^R} - \overline{\mathcal{P}_h^L}}{\overline{\mathcal{P}_h^R} + \overline{\mathcal{P}_h^L}}, \tag{62}$$

where again an overline denotes the time average.

In order to analyze the features of the SIGWs in this model, we consider the curvature perturbation with the $\delta$-function-type power spectrum [18,146]

$$\mathcal{P}_\zeta(k) = \mathcal{A}_\zeta\delta(\ln(k/k_p)). \tag{63}$$

After some straightforward calculations, we obtain the fractional energy density of the SIGWs

$$\Omega_{\mathrm{GW}}(k) = \frac{\mathcal{A}_\zeta^2\bar{k}^{-2}}{12}\left(\frac{4-\bar{k}^2}{4}\right)^2 \times \sum_{A=R,L}\frac{\overline{\tilde{I}^A(k,\bar{k}^{-1},\bar{k}^{-1},x\to\infty)^2}}{(z^A)^2}\Theta(2-\bar{k}), \tag{64}$$

where $\bar{k} = k/k_p$. The degree of circular polarization can be obtained easily, which is

$$\Pi = \frac{\mathcal{N}(k,\bar{k}^{-1},\bar{k}^{-1})}{\mathcal{M}(k,\bar{k}^{-1},\bar{k}^{-1})}\Theta(2-\bar{k}), \tag{65}$$

where

$$\mathcal{N}(k,u,v) = \frac{\overline{I^R(k,u,v,x\to\infty)^2}}{(z^R(k))^2} - \frac{\overline{I^L(k,u,v,x\to\infty)^2}}{(z^L(k))^2}, \tag{66}$$

and

$$\mathcal{M}(k,u,v) = \frac{\overline{I^R(k,u,v,x\to\infty)^2}}{(z^R(k))^2} + \frac{\overline{I^L(k,u,v,x\to\infty)^2}}{(z^L(k))^2}. \tag{67}$$

We compute the energy density and the degree of circular polarization induced by power spectrum (63) numerically and show the results in Fig. 1 through Fig. 4 for the two cases (44) and (45) that we mentioned in the above section.

### 5.1 The case $2\alpha\epsilon^S - 3 = 0$, $2\beta_{\mathcal{A}}\epsilon^S = 7$ and $2\mathcal{C}_1 + \mathcal{C}_3 = 0$

Combining the expression of $\overline{(I^A)^2}$ (B12) and Eqs. (64), (65), we plot the fractional energy density of the SIGWs from GR and chiral scalar–tensor theory of gravity in Fig. 1 and the left panel of Fig. 2, respectively.

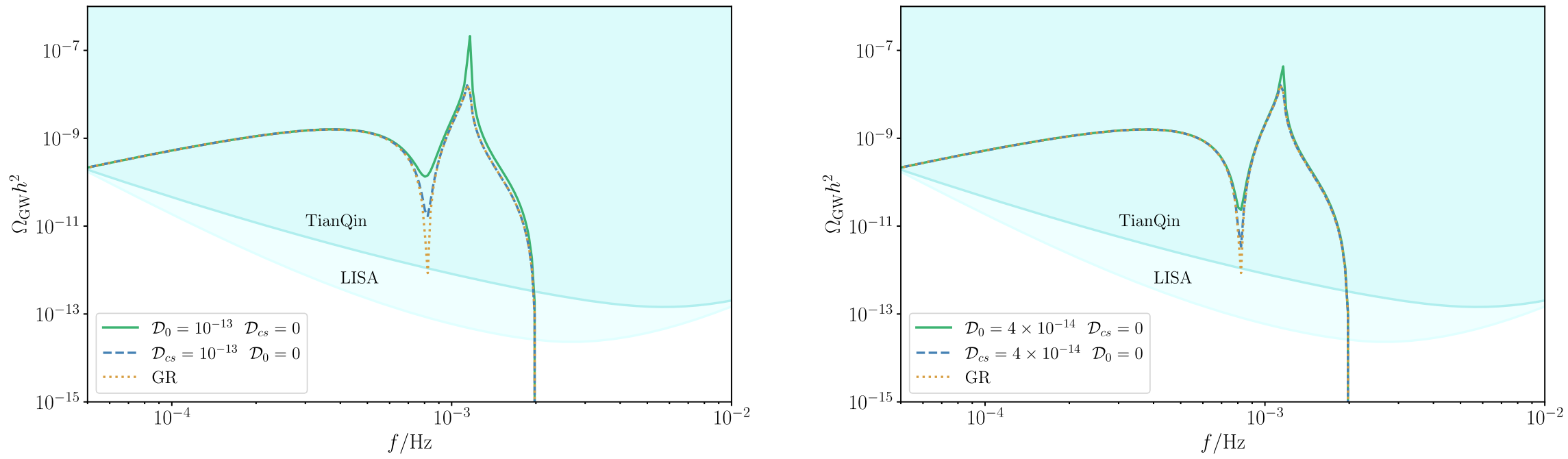

**Fig. 1** The energy density of SIGWs from GR and chiral scalar–tensor theory of gravity. The peak scale is $k_p = 10^{12}\mathrm{Mpc}^{-1}$, which corresponds to the maximum sensitivity of TianQin and LISA. The amplitude of the power spectrum is fixed to be $\mathcal{A}_\zeta = 10^{-2}$

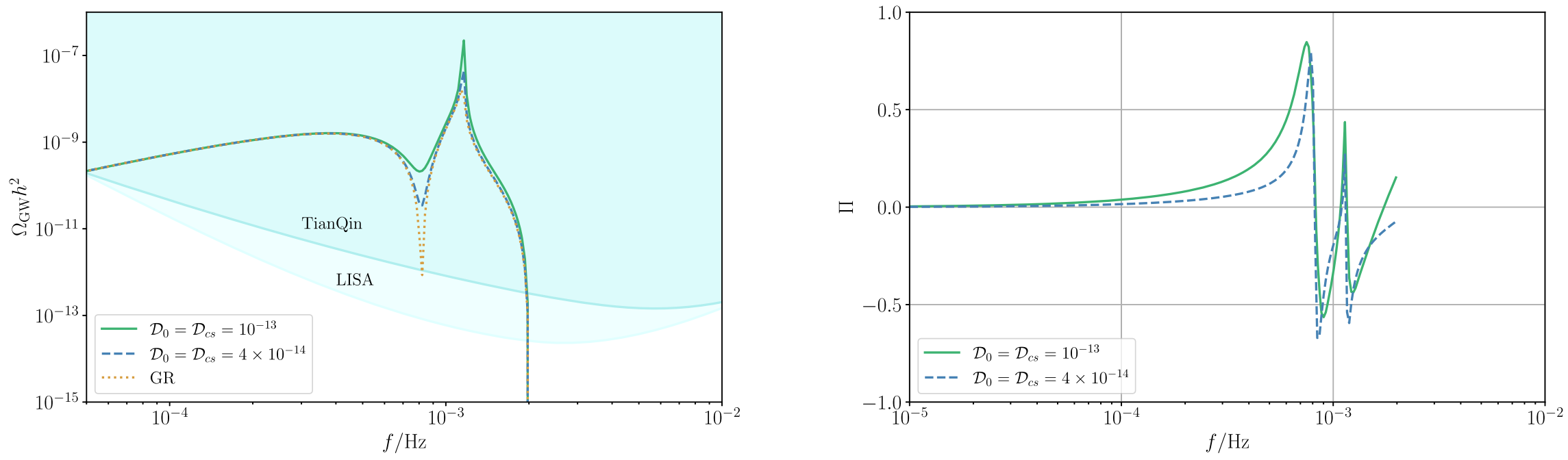

**Fig. 2** The energy density and degree of the circular polarization of SIGWs from GR and chiral scalar–tensor theory of gravity. The peak amplitude and the peak scale are fixed to be $\mathcal{A}_\zeta = 10^{-2}$ and $k_p = 10^{12}\mathrm{Mpc}^{-1}$, respectively

From Fig. 1, we observe that the contribution from $I^A_{\rm PV1}$ to the energy density $\Omega_{\rm GW}$ is larger than that from $I^A_{\rm CS}$, even if the parameters $\mathcal{O}(\mathcal{D}_{cs}) \sim \mathcal{O}(\mathcal{D}_0)$, mainly due to the terms $x^3 T^{***}_\psi T_\psi$ and $x^3 T^{**}_\psi T^*_\psi$ in $f^A_{\rm PV1}$ (53).

From the left panel of Fig. 2, we observe that with the increasing of $k$, the contribution from the PV terms to $\Omega_{\rm GW}$ approaches that of GR, but the constraint $(\mathcal{D}_{cs} + \mathcal{D}_0)k < 1$ effectively prevents the former from dominating. In the right panel of Fig. 2, it is obvious that the contribution of the PV terms to the degree of circular polarization $\Pi$ is negligible at lower frequencies. This can be attributed to the small values of the parameter $(\mathcal{D}_{cs} + \mathcal{D}_0)k \ll 1$ in the PV source term. However, as the frequency increases, the contribution from the PV terms to $\Omega_{\rm GW}$ gradually approaches that of GR, resulting in a significant enhancement in the degree of circular polarization $|\Pi|$.

### 5.2 The case $2\alpha\epsilon^S - 3 = 0$, $2\beta_{\mathcal{A}}\epsilon^S = 4$ and $2\mathcal{C}_1 + \mathcal{C}_3 = 0$

From Eqs. (B16), (B17) in Appendix B, we observe that the coefficient of $I^A_{\rm PV1}$ contains $\mathcal{D}_0 k^4 \lambda^A \eta_0^3$, which may result in different characteristic features in the fractional energy density of SIGWs. In this case, the time average of $(I^A)^2$ is

$$
\begin{aligned}
&\overline{I^A(k,u,v,x\to\infty)^2}\\
&= \frac{1}{2x^2}\Bigg\{\Bigg[\frac{3\left(u^2+v^2-3\right)+9\mathcal{D}_{cs}k\lambda^A}{4u^3v^3}\\
&\times\left(-4uv+\left(u^2+v^2-3\right)\ln\left|\frac{3-(u+v)^2}{3-(u-v)^2}\right|\right)\\
&+\frac{\widehat{\mathcal{D}}_0 k\lambda^A\left(u^2+v^2\right)}{270u^3v^3}\left(\frac{k}{k_p}\right)^3\\
&\times\Big(-(-27+45u^2+45v^2)\\
&+\left(15\sqrt{3}u^3+5\sqrt{3}u^3v^2-\sqrt{3}u^5\right)\\
&+\left(15\sqrt{3}v^3+5\sqrt{3}u^2v^3-\sqrt{3}v^5\right)\Big)\pi\Theta(u+v-\sqrt{3})\Bigg]^2\\
&+\Bigg[-\frac{3\left(u^2+v^2-3\right)^2+9\mathcal{D}_{cs}k\lambda^A\left(u^2+v^2-3\right)}{4u^3v^3}\\
&\pi\Theta(u+v-\sqrt{3})\\
&-\frac{\widehat{\mathcal{D}}_0 k\lambda^A\left(u^2+v^2\right)}{270u^3v^3}\left(\frac{k}{k_p}\right)^3\Big(-12\left(u^2+v^2+3\right)\\
&+(-27+45u^2+45v^2)\ln\left|\frac{3-(u+v)^2}{3-(u-v)^2}\right|
\end{aligned}
$$

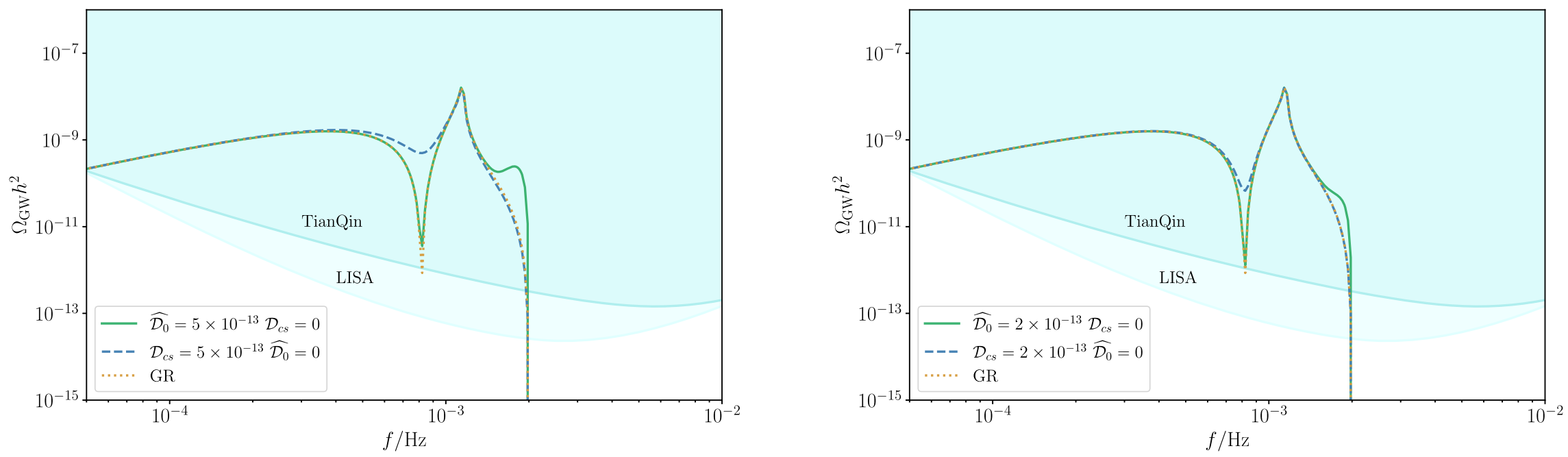

**Fig. 3** The energy density of SIGWs from GR and chiral scalar–tensor theory of gravity. The peak amplitude and the peak scale are fixed to be $\mathcal{A}_\zeta = 10^{-2}$ and $k_p = 10^{12}\text{Mpc}^{-1}$, respectively

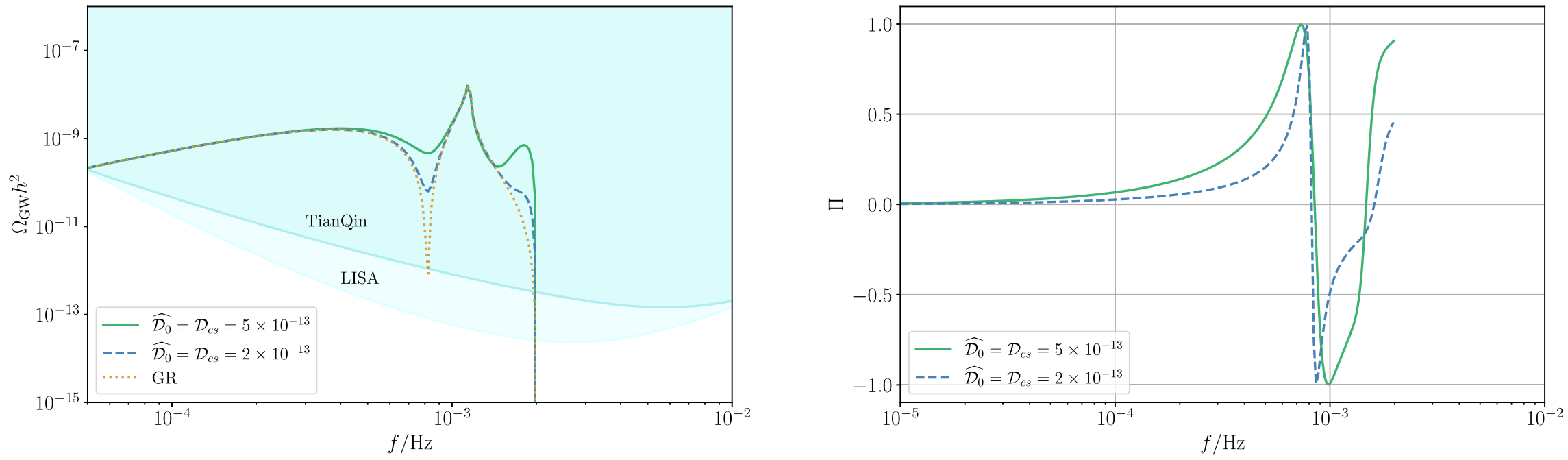

**Fig. 4** The energy density of SIGWs from GR and chiral scalar–tensor gravity. The peak amplitude and the peak scale are fixed to be $\mathcal{A}_\zeta = 10^{-2}$ and $k_p = 10^{12}\text{Mpc}^{-1}$, respectively

$$+ (15\sqrt{3}u^3 + 5\sqrt{3}u^3v^2 - \sqrt{3}u^5)\ln\left|\frac{u^2-(v+\sqrt{3})^2}{u^2-(v-\sqrt{3})}\right|$$
$$+ (15\sqrt{3}v^3 + 5\sqrt{3}u^2v^3 - \sqrt{3}v^5)\ln\left|\frac{v^2-(u+\sqrt{3})^2}{v^2-(u-\sqrt{3})}\right|\Bigg)\Bigg]^2\Bigg\}, \tag{68}$$

where $\widehat{\mathcal{D}}_0 = \mathcal{D}_0(k_p\eta_0)^3$ and $k_p$ is the peak scale. Plugging it into Eqs. (64), (65), we can compute the fractional energy density $\Omega_{\rm GW}$ and the degree of circular polarization $\Pi$ of the SIGWs, which are shown in Figs. 3 and 4.

From Fig. 3, we can observe two distinct regions, in which the SIGWs in chiral scalar–tensor gravity are obviously different from those in GR. In low frequencies, $f/f_p = k/k_p < 1$, from Eq. (68), the contribution from $I^A_{\rm PV1}$ is suppressed by a factor $(k/k_p)^3$ and is negligible. Nevertheless, in the high-frequency region, $k/k_p > 1$, the contribution from $I^A_{\rm PV1}$ dominates, resulting in a small peak protrusion.

In order to verify our above analysis on SIGWs, we also compute the fractional energy density $\Omega_{\rm GW}$ with $\mathcal{D}_{cs} = \widehat{\mathcal{D}}_0$, of which the results are shown in Fig. 4. With the increasing of frequency, the energy density of SIGWs from chiral scalar–tensor gravity deviates from that in GR, which results in a large degree of circular polarization $|\Pi|$, as seen in the right panel of Fig. 4.

## 6 Conclusion

In this paper, we investigate SIGWs in the chiral scalar–tensor theory of gravity. In our model, the PV terms consist of the CS term and the PV terms built from quadratic terms of the Riemann tensor and the first-order derivative of the scalar field. With the consideration of these PV terms, we expand the action up to the third order and focus on the quadratic and cubic actions relevant to SIGWs.

From the EOM for SIGWs (22), we can see that the amplitude and velocity birefringence effects, encoded in the parameter $\mu_A$, exist during the propagation of SIGWs. To solve the EOM for SIGWs during the radiation-dominated era, we analyze Green's function, which depends on $\mu_A$ and $z_A$. Since $\mu_A$ is relevant to the speed of GWs, we assume $\mu_A = 0$ because observations indicate that the deviation of the speed of GWs from that of light is very small. We find two special cases where $\mu_A = 0$ and $z^A$ are independent of $\eta$ with the exponential form of the coupling functions, in which we can obtain the analytical solution for the Green's function and further derive semi-analytic expressions to calculate SIGWs. One case occurs when $2\alpha\epsilon^S - 3 = 0$, $2\beta_\mathcal{A}\epsilon^S = 7$, and $2\mathcal{C}_1 + \mathcal{C}_3 = 0$. The other case occurs when $2\alpha\epsilon^S - 3 = 0$, $2\beta_\mathcal{A}\epsilon^S = 4$, and $2\mathcal{C}_1 + \mathcal{C}_3 = 0$. In both cases, the Green's function is the same as that in GR. However, the source terms are different and cause different characteristic features of SIGWs.

To analyze the features of SIGWs in chiral scalar–tensor gravity, we calculate the fractional energy density of SIGWs induced by the monochromatic power spectrum. By comparing the contributions from the CS term, $I^A_{\rm CS}$, and the PV scalar–tensor term, $I^A_{\rm PV1}$, at small scales, we analyze two specific cases mentioned before. In the first case, around the peak located at $\bar{f}=2/\sqrt{3}$, the contribution from $I^A_{\rm PV1}$ dominates over that from $I^A_{\rm CS}$. In the second case, near the region $\bar{f}<2/\sqrt{3}$, the term $I^A_{\rm CS}$ contributes more significantly than $I^A_{\rm PV1}$. However, near the region $\bar{f}>2/\sqrt{3}$, the contribution from $I^A_{\rm PV1}$ surpasses that from $I^A_{\rm CS}$, resulting in a small protrusion in the energy density. Additionally, we observe that the circular polarization degree of SIGWs can reach a significant level in both cases.

**Acknowledgements** Jia-Xi Feng would like to thank Prof. Tao Zhu for the useful discussion and Jin Qiao for providing help in the program.

**Funding** This work was partly supported by the National Natural Science Foundation of China (NSFC) under the grants No. 11975020 (XG) and No. 12305075 (FGZ).

**Data Availability Statement** This manuscript has no associated data or the data will not be deposited. [Authors' comment: This is a theoretical study and no experimental data.]

## Appendix A: Equations of motion

In this appendix, we give the EOMs of background and linear scalar perturbations as follows

$$3\mathcal{H}^2=\kappa^2a^2\bar{\rho}, \tag{A1}$$

$$-\mathcal{H}^2-2\mathcal{H}'=\kappa^2a^2\bar{P}, \tag{A2}$$

with

$$\bar{\rho}=\frac{1}{2a^2}\varphi'^2+V(\varphi),\quad \bar{P}=\frac{1}{2a^2}\varphi'^2-V(\varphi) \tag{A3}$$

where $\bar{\rho}$ and $\bar{P}$ represent the background energy density and pressure, respectively.

In the absence of the anisotropic stress, the EOMs of the linear scalar perturbations are

$$\nabla^2\psi-3\mathcal{H}(\psi'+\mathcal{H}\phi)=\frac{1}{2}\kappa^2(\varphi'\delta\varphi'-\varphi'^2\psi+a^2\,V_\varphi\delta\varphi), \tag{A4}$$

$$\psi'+\mathcal{H}\phi=\frac{1}{2}\kappa^2\varphi'\delta\varphi, \tag{A5}$$

$$\psi''+2\mathcal{H}\psi'+\mathcal{H}\phi'+(\mathcal{H}^2+2\mathcal{H}')\phi=\frac{1}{2}\kappa^2\Big(\varphi'\delta\varphi'-\varphi'^2\psi-a^2\,V_\varphi\delta\varphi\Big), \tag{A6}$$

$$\psi-\phi=0. \tag{A7}$$

From the above EOMs, the PV terms do not affect the evolution of background and linear scalar perturbations.

## Appendix B: The integral kernel

For the sake of clarity, we split $I^A$ defined in Eq. (58) into two parts as follows

$$I^A(k,u,v,x)=\frac{\sin(x)}{x}\Big(I_{\rm GRs}+I^A_{\rm CSs}+I^A_{\rm PV1s}\Big)+\frac{\cos(x)}{x}\Big(I_{\rm GRc}+I^A_{\rm CSc}+I^A_{\rm PV1c}\Big), \tag{B1}$$

where the subscript "s" and "c" stand for contributions involving the sine and cosine functions, respectively. And we can write

$$\begin{aligned} I^A_{\rm PV1s}&=\mathcal{I}_{\rm PV1s}(k,u,v,x)-\mathcal{I}_{\rm PV1s}(k,u,v,0),\\ I^A_{\rm PV1c}&=\mathcal{I}^A_{\rm PV1c}(k,u,v,x)-\mathcal{I}^A_{\rm PV1c}(k,u,v,0), \end{aligned} \tag{B2}$$

where $\mathcal{I}_{\rm PV1s}$ and $\mathcal{I}^A_{\rm PV1c}$ are defined by

$$\begin{aligned} \mathcal{I}_{\rm PV1s}(u,v,y)&=\int {\rm d}y\ \cos(y)y\cdot f_{\rm PV1}(k,u,v,y),\\ \mathcal{I}_{\rm PV1c}(u,v,y)&=-\int {\rm d}y\ \sin(y)y\cdot f_{\rm PV1}(k,u,v,y). \end{aligned} \tag{B3}$$

1. $2\alpha\epsilon^S-3=0$, $2\beta_{\mathcal{A}}\epsilon^S=7$ and $2\mathcal{C}_1+\mathcal{C}_3=0$

After tedious manipulations, the concrete expressions of $\mathcal{I}_{\rm PV1s}$ and $\mathcal{I}^A_{\rm PV1c}$ are found to be

$$\begin{aligned} \mathcal{I}_{\rm PV1s}=\frac{\mathcal{D}_0k\lambda^A}{4u^3v^3y^4}\Big\{&216uvy^2\cos y\cos\frac{uy}{\sqrt{3}}\cos\frac{vy}{\sqrt{3}}\\ &-72uvy^3\sin y\cos\frac{uy}{\sqrt{3}}\cos\frac{vy}{\sqrt{3}}\\ &+72\sqrt{3}vy^2\sin y\sin\frac{uy}{\sqrt{3}}\cos\frac{vy}{\sqrt{3}}\\ &+72\sqrt{3}uy^2\sin y\cos\frac{uy}{\sqrt{3}}\sin\frac{vy}{\sqrt{3}} \end{aligned}$$

$$+4\sqrt{3}vy\left[-54+\left(9-2v^2+4u^2\right)y^2\right]\cos y\sin\frac{uy}{\sqrt{3}}$$
$$\times\cos\frac{vy}{\sqrt{3}}+4\sqrt{3}uy\left[-54+\left(9-2u^2+4v^2\right)y^2\right]$$
$$\times\cos y\cos\frac{uy}{\sqrt{3}}\sin\frac{vy}{\sqrt{3}}-12\left[-54+\left(9+4u^2+4v^2\right)y^2\right]$$
$$\times\cos y\sin\frac{uy}{\sqrt{3}}\sin\frac{vy}{\sqrt{3}}-12y\left[18+\left(-9+2u^2\right.\right.$$
$$\left.\left.+2v^2\right)y^2\right]\sin y\sin\frac{uy}{\sqrt{3}}\sin\frac{vy}{\sqrt{3}}\Bigg\}$$
$$+\frac{\mathcal{D}_0k\lambda^A}{4u^3v^3}\Bigg\{\left(27-15u^2-15v^2-4u^2v^2-2u^4-2v^4\right)$$
$$\times\Bigg(\mathrm{Ci}\Big[\Big(\frac{u+v}{\sqrt{3}}+1\Big)y\Big]+\mathrm{Ci}\Big[\Big(\frac{u+v}{\sqrt{3}}-1\Big)y\Big]$$
$$-\mathrm{Ci}\Big[\Big(\frac{u-v}{\sqrt{3}}+1\Big)y\Big]-\mathrm{Ci}\Big[\Big|\Big(\frac{u-v}{\sqrt{3}}-1\Big)y\Big|\Big]\Bigg)\Bigg\}$$
$$+O_s(u,v,y), \tag{B4}$$

and

$$\mathcal{I}_{\mathrm{PV1c}}=\frac{\mathcal{D}_0k\lambda^A}{4u^3v^3y^4}\Bigg\{-216uvy^2\sin y\cos\frac{uy}{\sqrt{3}}\cos\frac{vy}{\sqrt{3}}$$
$$-72uvy^3\cos y\cos\frac{uy}{\sqrt{3}}\cos\frac{vy}{\sqrt{3}}$$
$$+72\sqrt{3}vy^2\cos y\sin\frac{uy}{\sqrt{3}}\cos\frac{vy}{\sqrt{3}}$$
$$+72\sqrt{3}uy^2\cos y\cos\frac{uy}{\sqrt{3}}\sin\frac{vy}{\sqrt{3}}$$
$$-4\sqrt{3}vy\left[-54+\left(9-2v^2+4u^2\right)y^2\right]$$
$$\times\sin y\sin\frac{uy}{\sqrt{3}}\cos\frac{vy}{\sqrt{3}}$$
$$-4\sqrt{3}uy\left[-54+\left(9-2u^2+4v^2\right)y^2\right]$$
$$\times\sin y\cos\frac{uy}{\sqrt{3}}\sin\frac{vy}{\sqrt{3}}$$
$$+12\left[-54+\left(9+4u^2+4v^2\right)y^2\right]\sin y\sin\frac{uy}{\sqrt{3}}\sin\frac{vy}{\sqrt{3}}$$
$$-12y\left[18+\left(-9+2u^2+2v^2\right)y^2\right]\cos y\sin\frac{uy}{\sqrt{3}}\sin\frac{vy}{\sqrt{3}}\Bigg\}$$
$$+\frac{\mathcal{D}_0k\lambda^A}{4u^3v^3}\Bigg\{-\left(27-15u^2-15v^2-4u^2v^2-2u^4-2v^4\right)$$
$$\times\Bigg(\mathrm{Si}\Big[\Big(1+\frac{u+v}{\sqrt{3}}\Big)y\Big]+\mathrm{Si}\Big[\Big(1-\frac{u+v}{\sqrt{3}}\Big)y\Big]$$
$$-\mathrm{Si}\Big[\Big(1+\frac{u-v}{\sqrt{3}}\Big)y\Big]-\mathrm{Si}\Big[\Big(1-\frac{u-v}{\sqrt{3}}\Big)y\Big]\Bigg)\Bigg\}$$
$$+O_c(u,v,y), \tag{B5}$$

where

$$\mathrm{Si}(x)=\int_0^x\mathrm{d}y\frac{\sin y}{y},\quad \mathrm{Ci}(x)=-\int_x^\infty\mathrm{d}y\frac{\cos y}{y}, \tag{B6}$$

$$O_s(u,v,y)=\frac{4\sqrt{3}\mathcal{D}_0k\lambda^A\left(u^2+v^2\right)}{3u^2v^2\left(u^4-2u^2\left(v^2+3\right)+\left(v^2-3\right)^2\right)}$$
$$\times\Bigg\{3u\left(u^2-v^2-3\right)\sin y\sin\frac{uy}{\sqrt{3}}\cos\frac{vy}{\sqrt{3}}$$
$$+3v\left(v^2-u^2-3\right)\sin y\cos\frac{uy}{\sqrt{3}}\sin\frac{vy}{\sqrt{3}}$$
$$+6\sqrt{3}uv\cos y\sin\frac{uy}{\sqrt{3}}\sin\frac{vy}{\sqrt{3}}$$
$$+\sqrt{3}(u^4+v^4-2u^2v^2-3u^2-3v^2)$$
$$\times\cos y\cos\frac{uy}{\sqrt{3}}\cos\frac{vy}{\sqrt{3}}\Bigg\}, \tag{B7}$$

and

$$O_c(u,v,y)$$
$$=-\frac{4\sqrt{3}\mathcal{D}_0k\lambda^A\left(u^2+v^2\right)}{3u^2v^2\left(u^4-2u^2\left(v^2+3\right)+\left(v^2-3\right)^2\right)}$$
$$\times\Bigg\{-3u\left(u^2-v^2-3\right)\cos y\sin\frac{uy}{\sqrt{3}}\cos\frac{vy}{\sqrt{3}}$$
$$-3v\left(v^2-u^2-3\right)\cos y\cos\frac{uy}{\sqrt{3}}\sin\frac{vy}{\sqrt{3}}$$
$$+6\sqrt{3}uv\sin y\sin\frac{uy}{\sqrt{3}}\sin\frac{vy}{\sqrt{3}}$$
$$+\sqrt{3}\left(u^4+v^4-2u^2v^2-3u^2-3v^2\right)$$
$$\times\sin y\cos\frac{uy}{\sqrt{3}}\cos\frac{vy}{\sqrt{3}}\Bigg\}. \tag{B8}$$

Since we are interested in the SIGWs at the present time, we take $x\gg 1$. In this limit, we have

$$I^A_{\mathrm{PV1}s}(k,u,v,x)\Big|_{x\to\infty}$$
$$=\frac{\mathcal{D}_0k\lambda^A}{4u^3v^3}\Big[-4uv(9-2u^2-2v^2)$$
$$+(2u^4+2v^4+4u^2v^2+15u^2+15v^2-27)$$
$$\times\ln\left|\frac{3-(u+v)^2}{3-(u-v)^2}\right|\Big]+O_s(u,v,x\to\infty)$$
$$-\frac{4\mathcal{D}_0k\lambda\left(u^2+v^2\right)\left(u^4+v^4-2u^2v^2-3u^2-3v^2\right)}{u^2v^2\left(u^4-2u^2\left(v^2+3\right)+\left(v^2-3\right)^2\right)}, \tag{B9}$$

where the last term corresponds to $O_s(u,v,x\to 0)$, which results in obvious deviation with CS gravity. And

$$I^A_{\mathrm{PV1}c}(k,u,v,x)\Big|_{x\to\infty}$$
$$=-\frac{\mathcal{D}_0k\lambda^A}{4u^3v^3}\Big[(2u^4+2v^4+4u^2v^2+15u^2$$
$$+15v^2-27)\pi\Theta(u+v-\sqrt{3})\Big]+O_c(u,v,x\to\infty). \tag{B10}$$

Similarly, for the other terms, we have

$$
\begin{aligned}
&I_{\rm GRs}(k,u,v,x)\Big|_{x\to\infty} \\
&= \frac{3\left(u^2+v^2-3\right)}{4u^3v^3}\left(-4uv+\left(u^2+v^2-3\right)\right. \\
&\quad\left.\times\ln\left|\frac{3-(u+v)^2}{3-(u-v)^2}\right|\right), \\
&I_{\rm GRc}(k,u,v,x)\Big|_{x\to\infty} \\
&= -\frac{3\left(u^2+v^2-3\right)^2}{4u^3v^3}\pi\Theta(u+v-\sqrt{3}), \\
&I_{\rm CSs}(k,u,v,x)\Big|_{x\to\infty} \\
&= \frac{9\mathcal{D}_{cs}k\lambda^A}{4u^3v^3}\left(-4uv+\left(u^2+v^2-3\right)\ln\left|\frac{3-(u+v)^2}{3-(u-v)^2}\right|\right), \\
&I_{\rm CSc}(k,u,v,x)\Big|_{x\to\infty} \\
&= -\frac{9\mathcal{D}_{cs}k\lambda^A}{4u^3v^3}\pi\left(u^2+v^2-3\right)\Theta(u+v-\sqrt{3}). \qquad \text{(B11)}
\end{aligned}
$$

As a result, the time average is

$$
\begin{aligned}
&\overline{I^A(k,u,v,x\to\infty)^2} \\
&= \frac{1}{2x^2}\left\{\Big[I_{{\rm GR}s}(u,v,x)+I^A_{{\rm CS}s}(k,u,v,x)\right. \\
&\quad+I^A_{{\rm PV1}s}(k,u,v,x)-O_s(u,v,x)\Big]^2_{x\to\infty} \\
&\quad+\Big[I_{{\rm GR}c}(u,v,x)+I^A_{{\rm CS}c}(k,u,v,x) \\
&\quad\left.+I^A_{{\rm PV1}c}(k,u,v,x)-O_c(u,v,x)\Big]^2_{x\to\infty}+\overline{O}\right\}. \qquad \text{(B12)}
\end{aligned}
$$

where

$$
\begin{aligned}
\overline{O} &= 24\left(\frac{\mathcal{D}_0k\lambda^A\left(u^2+v^2\right)}{u^2v^2\left(u^4-2u^2\left(v^2+3\right)+\left(v^2-3\right)^2\right)}\right)^2 \\
&\quad\times\left[u^2\left(u^2-v^2-3\right)^2+v^2\left(v^2-u^2-3\right)^2\right]. \qquad \text{(B13)}
\end{aligned}
$$

2. $2\alpha\epsilon^S-3=0$, $2\beta_{\mathcal{A}}\epsilon^S=4$, and $2\mathcal{C}_1+\mathcal{C}_3=0$

In this case, $\mathcal{I}_{\rm PV1s}$ and $\mathcal{I}^A_{\rm PV1c}$ are

$$
\begin{aligned}
\mathcal{I}_{\rm PV1s} &= \frac{\mathcal{D}_0k\lambda^A}{45u^3v^3y^5}\left(u^2+v^2\right)k^3\eta_0^3 \\
&\quad\times\left\{-36uvy^3\sin y\cos\frac{uy}{\sqrt{3}}\cos\frac{vy}{\sqrt{3}}\right. \\
&\quad+2uvy^2\left(72+(-6+u^2+v^2)y^2\right) \\
&\quad\times\cos y\cos\frac{uy}{\sqrt{3}}\cos\frac{vy}{\sqrt{3}} \\
&\quad-2\sqrt{3}vy\left(72+(-6+v^2-2u^2)y^2\right) \\
&\quad\times\cos y\sin\frac{uy}{\sqrt{3}}\cos\frac{vy}{\sqrt{3}} \\
&\quad-2\sqrt{3}uy\left(72+(-6+u^2-2v^2)y^2\right) \\
&\quad\times\cos y\cos\frac{uy}{\sqrt{3}}\sin\frac{vy}{\sqrt{3}} \\
&\quad-2\sqrt{3}vy^2\left(-18+\left(3+v^2-2u^2\right)y^2\right) \\
&\quad\times\sin y\sin\frac{uy}{\sqrt{3}}\cos\frac{vy}{\sqrt{3}} \\
&\quad-2\sqrt{3}uy^2\left(-18+\left(3+u^2-2v^2\right)y^2\right) \\
&\quad\times\sin y\cos\frac{uy}{\sqrt{3}}\sin\frac{vy}{\sqrt{3}} \\
&\quad+2\Big[216-6\left(u^2+v^2+3\right)y^2 \\
&\quad+\left(9+u^4+v^4-4u^2v^2-12u^2-12v^2\right)y^4\Big] \\
&\quad\times\cos y\sin\frac{uy}{\sqrt{3}}\sin\frac{vy}{\sqrt{3}}-6y\left(18+(2u^2+2v^2-3)y^2\right) \\
&\quad\left.\times\sin y\sin\frac{uy}{\sqrt{3}}\sin\frac{vy}{\sqrt{3}}\right\} \\
&\quad+\frac{\mathcal{D}_0k\lambda^A}{45u^3v^3}\frac{\left(u^2+v^2\right)k^3\eta_0^3}{6}\left\{(-27+45u^2+45v^2)\right. \\
&\quad\times\left(\mathrm{Si}\Big[\Big(1+\frac{u+v}{\sqrt{3}}\Big)y\Big]+\mathrm{Si}\Big[\Big(1-\frac{u+v}{\sqrt{3}}\Big)y\Big]\right. \\
&\quad\left.-\mathrm{Si}\Big[\Big(1+\frac{u-v}{\sqrt{3}}\Big)y\Big]-\mathrm{Si}\Big[\Big(1-\frac{u-v}{\sqrt{3}}\Big)y\Big]\right) \\
&\quad+(15\sqrt{3}u^3+5\sqrt{3}u^3v^2-\sqrt{3}u^5)\left(\mathrm{Si}\Big[\Big(1+\frac{u+v}{\sqrt{3}}\Big)y\Big]\right. \\
&\quad-\mathrm{Si}\Big[\Big(1-\frac{u+v}{\sqrt{3}}\Big)y\Big]-\mathrm{Si}\Big[\Big(1+\frac{u-v}{\sqrt{3}}\Big)y\Big] \\
&\quad\left.+\mathrm{Si}\Big[\Big(1-\frac{u-v}{\sqrt{3}}\Big)y\Big]\right) \\
&\quad+(15\sqrt{3}v^3+5\sqrt{3}u^2v^3-\sqrt{3}v^5)\left(\mathrm{Si}\Big[\Big(1+\frac{u+v}{\sqrt{3}}\Big)y\Big]\right. \\
&\quad-\mathrm{Si}\Big[\Big(1-\frac{u+v}{\sqrt{3}}\Big)y\Big]+\mathrm{Si}\Big[\Big(1+\frac{u-v}{\sqrt{3}}\Big)y\Big] \\
&\quad\left.\left.-\mathrm{Si}\Big[\Big(1-\frac{u-v}{\sqrt{3}}\Big)y\Big]\right)\right\}, \qquad \text{(B14)}
\end{aligned}
$$

$$
\begin{aligned}
\mathcal{I}_{\rm PV1c} &= \frac{\mathcal{D}_0k\lambda^A}{45u^3v^3y^5}\left(u^2+v^2\right)k^3\eta_0^3\left\{-36uvy^3\cos y\cos\frac{uy}{\sqrt{3}}\right. \\
&\quad\times\cos\frac{vy}{\sqrt{3}}-2uvy^2\left(72+\left(-6+u^2+v^2\right)y^2\right)\sin y \\
&\quad\times\cos\frac{uy}{\sqrt{3}}\cos\frac{vy}{\sqrt{3}}+2\sqrt{3}vy\left(72+\left(-6+v^2-2u^2\right)y^2\right) \\
&\quad\times\sin y\sin\frac{uy}{\sqrt{3}}\cos\frac{vy}{\sqrt{3}}+2\sqrt{3}uy\left(72+\left(-6+u^2\right.\right. \\
&\quad\left.\left.-2v^2\right)y^2\right)\sin y\cos\frac{uy}{\sqrt{3}}\sin\frac{vy}{\sqrt{3}}-2\sqrt{3}vy^2 \\
&\quad\times\left(-18+\left(3+v^2-2u^2\right)y^2\right)\cos y\sin\frac{uy}{\sqrt{3}}\cos\frac{vy}{\sqrt{3}} \\
&\quad-2\sqrt{3}uy^2\left(-18+\left(3+u^2-2v^2\right)y^2\right)\cos y
\end{aligned}
$$

$$\times \cos\frac{uy}{\sqrt{3}}\sin\frac{vy}{\sqrt{3}} - 2\Big[216 - 6(u^2+v^2+3)y^2$$
$$+ (9 + u^4 + v^4 - 4u^2v^2 - 12u^2 - 12v^2)y^4\Big]$$
$$\times \sin y \sin\frac{uy}{\sqrt{3}}\sin\frac{vy}{\sqrt{3}} - 6y$$
$$\times \left(18 + \left(2u^2+2v^2-3\right)y^2\right)\cos y \sin\frac{uy}{\sqrt{3}}\sin\frac{vy}{\sqrt{3}}\Bigg\}$$
$$+ \frac{\mathcal{D}_0 k\lambda^A}{45u^3v^3}\frac{k^3\eta_0^3\left(u^2+v^2\right)}{6}$$
$$\times \Bigg\{(-27+45u^2+45v^2)\left(\mathrm{Ci}\Big[\Big(\frac{u+v}{\sqrt{3}}+1\Big)y\Big]\right.$$
$$+\mathrm{Ci}\Big[\Big(\frac{u+v}{\sqrt{3}}-1\Big)y\Big] - \mathrm{Ci}\Big[\Big(\frac{u-v}{\sqrt{3}}+1\Big)y\Big]$$
$$\left.-\mathrm{Ci}\Big[\Big|\Big(\frac{u-v}{\sqrt{3}}-1\Big)y\Big|\Big]\right)$$
$$+ (15\sqrt{3}u^3 + 5\sqrt{3}u^3v^2 - \sqrt{3}u^5)\left(\mathrm{Ci}\Big[\Big(\frac{u+v}{\sqrt{3}}+1\Big)y\Big]\right.$$
$$-\mathrm{Ci}\Big[\Big(\frac{u+v}{\sqrt{3}}-1\Big)y\Big] - \mathrm{Ci}\Big[\Big(\frac{u-v}{\sqrt{3}}+1\Big)y\Big]$$
$$\left.+\mathrm{Ci}\Big[\Big|\Big(\frac{u-v}{\sqrt{3}}-1\Big)y\Big|\Big]\right)$$
$$+ (15\sqrt{3}v^3 + 5\sqrt{3}u^2v^3 - \sqrt{3}v^5)\left(\mathrm{Ci}\Big[\Big(\frac{u+v}{\sqrt{3}}+1\Big)y\Big]\right.$$
$$-\mathrm{Ci}\Big[\Big(\frac{u+v}{\sqrt{3}}-1\Big)y\Big] + \mathrm{Ci}\Big[\Big(\frac{u-v}{\sqrt{3}}+1\Big)y\Big]$$
$$\left.-\mathrm{Ci}\Big[\Big|\Big(\frac{u-v}{\sqrt{3}}-1\Big)y\Big|\Big]\right)\Bigg\}. \tag{B15}$$

In the limit $x \gg 1$, we obtain

$$I^A_{\mathrm{PV1}s}(k,u,v,x)\Big|_{x\to\infty}$$
$$= \frac{\mathcal{D}_0 k\lambda^A\left(u^2+v^2\right)}{45u^3v^3}\left(\frac{k^3\eta_0^3}{6}\right)\Big(-\left(-27+45u^2+45v^2\right)$$
$$+ (15\sqrt{3}u^3 + 5\sqrt{3}u^3v^2 - \sqrt{3}u^5)$$
$$+ (15\sqrt{3}v^3 + 5\sqrt{3}u^2v^3 - \sqrt{3}v^5)\Big)\pi\Theta(u+v-\sqrt{3}), \tag{B16}$$

and

$$I^A_{\mathrm{PV1}c}(k,u,v,x)\Big|_{x\to\infty}$$
$$= -\frac{\mathcal{D}_0 k\lambda^A\left(u^2+v^2\right)}{45u^3v^3}\left(\frac{k^3\eta_0^3}{6}\right)\Big(-12\left(u^2+v^2+3\right)$$
$$+ \left(-27+45u^2+45v^2\right)\ln\left|\frac{3-(u+v)^2}{3-(u-v)^2}\right|$$
$$+ (15\sqrt{3}u^3 + 5\sqrt{3}u^3v^2 - \sqrt{3}u^5)\ln\left|\frac{u^2-(v+\sqrt{3})^2}{u^2-(v-\sqrt{3})}\right|$$
$$+ (15\sqrt{3}v^3 + 5\sqrt{3}u^2v^3 - \sqrt{3}v^5)\ln\left|\frac{v^2-(u+\sqrt{3})^2}{v^2-(u-\sqrt{3})}\right|\Big). \tag{B17}$$

As a result, the time averaged integral kernel is

$$\overline{I^A(k,u,v,x\to\infty)^2}$$
$$= \frac{1}{2x^2}\Bigg\{\Big[I_{\mathrm{GR}s}(u,v,x\to\infty) + I^A_{\mathrm{CS}s}(k,u,v,x\to\infty)$$
$$+ I^A_{\mathrm{PV1}s}(k,u,v,x\to\infty)\Big]^2$$
$$+ \Big[I_{\mathrm{GR}c}(u,v,x\to\infty) + I^A_{\mathrm{CS}c}(k,u,v,x\to\infty)$$
$$+ I^A_{\mathrm{PV1}c}(k,u,v,x\to\infty)\Big]^2\Bigg\}. \tag{B18}$$

## References

1. B.P. Abbott et al. (LIGO Scientific, Virgo), Phys. Rev. Lett. **116**, 131102 (2016). https://doi.org/10.1103/PhysRevLett.116.131102. arXiv:1602.03847 [gr-qc]
2. B.P. Abbott et al. (LIGO Scientific, Virgo), Phys. Rev. Lett. **116**, 241103 (2016). https://doi.org/10.1103/PhysRevLett.116.241103. arXiv:1606.04855 [gr-qc]
3. B.P. Abbott et al. (LIGO Scientific, Virgo), Phys. Rev. Lett. **116**, 061102 (2016). https://doi.org/10.1103/PhysRevLett.116.061102. arXiv:1602.03837 [gr-qc]
4. B.P. Abbott et al. (LIGO Scientific, Virgo), Astrophys. J. Lett. **851**, L35 (2017). https://doi.org/10.3847/2041-8213/aa9f0c. arXiv:1711.05578 [astro-ph.HE]
5. B.P. Abbott et al. (LIGO Scientific, Virgo), Phys. Rev. Lett. **119**, 161101 (2017). https://doi.org/10.1103/PhysRevLett.119.161101. arXiv:1710.05832 [gr-qc]
6. B.P. Abbott et al. (LIGO Scientific, Virgo), Phys. Rev. Lett. **119**, 141101 (2017). https://doi.org/10.1103/PhysRevLett.119.141101. arXiv:1709.09660 [gr-qc]
7. B.P. Abbott et al. (LIGO Scientific, VIRGO), Phys. Rev. Lett. **118**, 221101 (2017) [Erratum: Phys.Rev.Lett. 121, 129901 (2018)]. https://doi.org/10.1103/PhysRevLett.118.221101 arXiv:1706.01812 [gr-qc]
8. B.P. Abbott et al. (LIGO Scientific, Virgo), Phys. Rev. X **9**, 031040 (2019). https://doi.org/10.1103/PhysRevX.9.031040. arXiv:1811.12907 [astro-ph.HE]
9. R. Abbott et al. (LIGO Scientific, Virgo), Astrophys. J. Lett. **896**, L44 (2020). https://doi.org/10.3847/2041-8213/ab960f. arXiv:2006.12611 [astro-ph.HE]
10. B.P. Abbott et al. (LIGO Scientific, Virgo), Astrophys. J. Lett. **892**, L3 (2020). https://doi.org/10.3847/2041-8213/ab75f5. arXiv:2001.01761 [astro-ph.HE]
11. R. Abbott et al. (LIGO Scientific, Virgo), Phys. Rev. D **102**, 043015 (2020). https://doi.org/10.1103/PhysRevD.102.043015. arXiv:2004.08342 [astro-ph.HE]
12. C. Caprini, D.G. Figueroa, Class. Quantum Gravity **35**, 163001 (2018). https://doi.org/10.1088/1361-6382/aac608. arXiv:1801.04268 [astro-ph.CO]
13. D.G. Figueroa, E.H. Tanin, JCAP **08**, 011 (2019). https://doi.org/10.1088/1475-7516/2019/08/011. arXiv:1905.11960 [astro-ph.CO]
14. E.H. Tanin, T. Tenkanen, JCAP **01**, 053 (2021). https://doi.org/10.1088/1475-7516/2021/01/053. arXiv:2004.10702 [astro-ph.CO]

15. I.V. Fomin, S.V. Chervon, A.N. Morozov, I.S. Golyak, Eur. Phys. J. C **82**, 642 (2022). https://doi.org/10.1140/epjc/s10052-022-10601-9
16. S. Matarrese, O. Pantano, D. Saez, Phys. Rev. D **47**, 1311 (1993). https://doi.org/10.1103/PhysRevD.47.1311
17. S. Matarrese, O. Pantano, D. Saez, Phys. Rev. Lett. **72**, 320 (1994). https://doi.org/10.1103/PhysRevLett.72.320. arXiv:astro-ph/9310036
18. K.N. Ananda, C. Clarkson, D. Wands, Phys. Rev. D **75**, 123518 (2007). https://doi.org/10.1103/PhysRevD.75.123518. arXiv:gr-qc/0612013
19. D. Baumann, P.J. Steinhardt, K. Takahashi, K. Ichiki, Phys. Rev. D **76**, 084019 (2007). https://doi.org/10.1103/PhysRevD.76.084019. arXiv:hep-th/0703290
20. R. Saito, J. Yokoyama, Phys. Rev. Lett. **102**, 161101 (2009) [Erratum: Phys.Rev.Lett. 107, 069901 (2011)]. https://doi.org/10.1103/PhysRevLett.102.161101. arXiv:0812.4339 [astro-ph]
21. N. Orlofsky, A. Pierce, J.D. Wells, Phys. Rev. D **95**, 063518 (2017). https://doi.org/10.1103/PhysRevD.95.063518. arXiv:1612.05279 [astro-ph.CO]
22. T. Nakama, J. Silk, M. Kamionkowski, Phys. Rev. D **95**, 043511 (2017). https://doi.org/10.1103/PhysRevD.95.043511. arXiv:1612.06264 [astro-ph.CO]
23. K. Kohri, T. Terada, Phys. Rev. D **97**, 123532 (2018). https://doi.org/10.1103/PhysRevD.97.123532. arXiv:1804.08577 [gr-qc]
24. S. Kuroyanagi, T. Chiba, T. Takahashi, JCAP **11**, 038 (2018). https://doi.org/10.1088/1475-7516/2018/11/038. arXiv:1807.00786 [astro-ph.CO]
25. Y. Lu, A. Ali, Y. Gong, J. Lin, F. Zhang, Phys. Rev. D **102**, 083503 (2020). https://doi.org/10.1103/PhysRevD.102.083503(2020). arXiv:2006.03450 [gr-qc]
26. M. Braglia, X. Chen, D.K. Hazra, JCAP **03**, 005 (2021). https://doi.org/10.1088/1475-7516/2021/03/005. arXiv:2012.05821 [astro-ph.CO]
27. A. Romero-Rodriguez, M. Martinez, O. Pujolàs, M. Sakellariadou, V. Vaskonen, Phys. Rev. Lett. **128**, 051301 (2022). https://doi.org/10.1103/PhysRevLett.128.051301. arXiv:2107.11660 [gr-qc]
28. A. Ota, H.J. Macpherson, W.R. Coulton, Phys. Rev. D **106**, 063521 (2022). https://doi.org/10.1103/PhysRevD.106.063521. arXiv:2111.09163 [gr-qc]
29. J.-Z. Zhou, X. Zhang, Q.-H. Zhu, Z. Chang, JCAP **05**, 013 (2022). https://doi.org/10.1088/1475-7516/2022/05/013. arXiv:2106.01641 [astro-ph.CO]
30. G. Domènech, Universe **7**, 398 (2021). https://doi.org/10.3390/universe7110398. arXiv:2109.01398 [gr-qc]
31. K. Inomata, Phys. Rev. D **104**, 123525 (2021). https://doi.org/10.1103/PhysRevD.104.123525. arXiv:2109.06192 [astro-ph.CO]
32. C. Yuan, Q.-G. Huang, (2021). arXiv:2103.04739 [astro-ph.GA]
33. M. Sipp, B.M. Schaefer, Phys. Rev. D **107**, 063538 (2023). https://doi.org/10.1103/PhysRevD.107.063538. arXiv:2212.01190 [astro-ph.CO]
34. X. Zhang, J.-Z. Zhou, Z. Chang, Eur. Phys. J. C **82**, 781 (2022). https://doi.org/10.1140/epjc/s10052-022-10742-x. arXiv:2208.12948 [astro-ph.CO]
35. Z. Chang, X. Zhang, J.-Z. Zhou, Phys. Rev. D **107**, 063510 (2023). https://doi.org/10.1103/PhysRevD.107.063510. arXiv:2209.07693 [astro-ph.CO]
36. S. Wang, Z.-C. Zhao, Q.-H. Zhu, Phys. Rev. Res. **6**, 013207 (2024). https://doi.org/10.1103/PhysRevResearch.6.013207. arXiv:2307.03095 [astro-ph.CO]
37. S. Wang, Y.-F. Wang, Q.-G. Huang, T.G.F. Li, Phys. Rev. Lett. **120**, 191102 (2018). https://doi.org/10.1103/PhysRevLett.120.191102. arXiv:1610.08725 [astro-ph.CO]
38. S. Wang, V. Vardanyan, K. Kohri, Phys. Rev. D **106**, 123511 (2022). https://doi.org/10.1103/PhysRevD.106.123511. arXiv:2107.01935 [gr-qc]
39. Z. Yi, Q. Fei, (2022). arXiv:2210.03641 [astro-ph.CO]
40. T. Papanikolaou, C. Tzerefos, S. Basilakos, E.N. Saridakis, JCAP **10**, 013 (2022). https://doi.org/10.1088/1475-7516/2022/10/013. arXiv:2112.15059 [astro-ph.CO]
41. Z.-C. Chen, C. Yuan, Q.-G. Huang, Phys. Lett. B **829**, 137040 (2022). https://doi.org/10.1016/j.physletb.2022.137040. arXiv:2108.11740 [astro-ph.CO]
42. J. Kozaczuk, T. Lin, E. Villarama, Phys. Rev. D **105**, 123023 (2022). https://doi.org/10.1103/PhysRevD.105.123023. arXiv:2108.12475 [astro-ph.CO]
43. Z. Yi, JCAP **03**, 048 (2023). https://doi.org/10.1088/1475-7516/2023/03/048. arXiv:2206.01039 [gr-qc]
44. J. Cang, Y.-Z. Ma, Y. Gao, Astrophys. J. **949**, 64 (2023). https://doi.org/10.3847/1538-4357/acc949. arXiv:2210.03476 [astro-ph.CO]
45. Z.-C. Zhao, S. Wang, Universe **9**, 157 (2023). https://doi.org/10.3390/universe9040157. arXiv:2211.09450 [astro-ph.CO]
46. T. Papanikolaou, C. Tzerefos, S. Basilakos, E.N. Saridakis, Eur. Phys. J. C **83**, 31 (2023). https://doi.org/10.1140/epjc/s10052-022-11157-4. arXiv:2205.06094 [gr-qc]
47. Z. Chang, Y.-T. Kuang, X. Zhang, J.-Z. Zhou, Chin. Phys. C **47**, 055104 (2023). https://doi.org/10.1088/1674-1137/acc649. arXiv:2209.12404 [astro-ph.CO]
48. A. Ghoshal, A. Moursy, Q. Shafi, Phys. Rev. D **108**, 055039 (2023). https://doi.org/10.1103/PhysRevD.108.055039. arXiv:2306.04002 [hep-ph]
49. S. Balaji, G. Domènech, G. Franciolini, JCAP **10**, 041 (2023). https://doi.org/10.1088/1475-7516/2023/10/041. arXiv:2307.08552 [gr-qc]
50. J. Fumagalli, S. Renaux-Petel, L.T. Witkowski, JCAP **08**, 030 (2021). https://doi.org/10.1088/1475-7516/2021/08/030. arXiv:2012.02761 [astro-ph.CO]
51. J. Lin, Q. Gao, Y. Gong, Y. Lu, C. Zhang, F. Zhang, Phys. Rev. D **101**, 103515 (2020). https://doi.org/10.1103/PhysRevD.101.103515. arXiv:2001.05909 [gr-qc]
52. W. Ahmed, M. Junaid, U. Zubair, Nucl. Phys. B **984**, 115968 (2022). https://doi.org/10.1016/j.nuclphysb.2022.115968. arXiv:2109.14838 [astro-ph.CO]
53. F. Zhang, Phys. Rev. D **105**, 063539 (2022). https://doi.org/10.1103/PhysRevD.105.063539. arXiv:2112.10516 [gr-qc]
54. M. Solbi, K. Karami, Eur. Phys. J. C **81**, 884 (2021). https://doi.org/10.1140/epjc/s10052-021-09690-9. arXiv:2106.02863 [astro-ph.CO]
55. R.-G. Cai, C. Chen, C. Fu, Phys. Rev. D **104**, 083537 (2021). https://doi.org/10.1103/PhysRevD.104.083537. arXiv:2108.03422 [astro-ph.CO]
56. K. Rezazadeh, Z. Teimoori, S. Karimi, K. Karami, Eur. Phys. J. C **82**, 758 (2022). https://doi.org/10.1140/epjc/s10052-022-10735-w. arXiv:2110.01482 [gr-qc]
57. S. Balaji, J. Silk, Y.-P. Wu, JCAP **06**, 008 (2022). https://doi.org/10.1088/1475-7516/2022/06/008. arXiv:2202.00700 [astro-ph.CO]
58. R. Arya, A.K. Mishra, Phys. Dark Univ. **37**, 101116 (2022). https://doi.org/10.1016/j.dark.2022.101116. arXiv:2204.02896 [astro-ph.CO]
59. Y. Aldabergenov, A. Addazi, S.V. Ketov, Eur. Phys. J. C **82**, 681 (2022). https://doi.org/10.1140/epjc/s10052-022-10654-w. arXiv:2206.02601 [astro-ph.CO]
60. C. Fu, C. Chen, (2022). arXiv:2211.11387 [astro-ph.CO]
61. J.-X. Zhao, X.-H. Liu, N. Li, Phys. Rev. D **107**, 043515 (2023). https://doi.org/10.1103/PhysRevD.107.043515. arXiv:2302.06886 [astro-ph.CO]

62. B.-M. Gu, F.-W. Shu, K. Yang, (2023). arXiv:2307.00510 [astro-ph.CO]
63. S. Choudhury, A. Karde, S. Panda, M. Sami, (2023). arXiv:2308.09273 [astro-ph.CO]
64. G. Bhattacharya, S. Choudhury, K. Dey, S. Ghosh, A. Karde, N.S. Mishra, (2023). arXiv:2309.00973 [astro-ph.CO]
65. S. Choudhury, K. Dey, A. Karde, (2023). arXiv:2311.15065 [astro-ph.CO]
66. S. Choudhury, A. Karde, S. Panda, M. Sami, (2024). arXiv:2401.10925 [astro-ph.CO]
67. G. Domènech, Int. J. Mod. Phys. D **29**, 2050028 (2020). https://doi.org/10.1142/S0218271820500285. arXiv:1912.05583 [gr-qc]
68. G. Domènech, S. Pi, M. Sasaki, JCAP **08**, 017 (2020). arXiv:2005.12314 [gr-qc]
69. L.T. Witkowski, G. Domènech, J. Fumagalli, S. Renaux-Petel, JCAP **05**, 028 (2022). https://doi.org/10.1088/1475-7516/2022/05/028. arXiv:2110.09480 [astro-ph.CO]
70. R.-G. Cai, S. Pi, M. Sasaki, Phys. Rev. Lett. **122**, 201101 (2019). https://doi.org/10.1103/PhysRevLett.122.201101. arXiv:1810.11000 [astro-ph.CO]
71. J.R. Espinosa, D. Racco, A. Riotto, JCAP **09**, 012 (2018). https://doi.org/10.1088/1475-7516/2018/09/012. arXiv:1804.07732 [hep-ph]
72. P. Adshead, K.D. Lozanov, Z.J. Weiner, JCAP **10**, 080 (2021). https://doi.org/10.1088/1475-7516/2021/10/080. arXiv:2105.01659 [astro-ph.CO]
73. M.W. Davies, P. Carrilho, D.J. Mulryne, JCAP **06**, 019 (2022). https://doi.org/10.1088/1475-7516/2022/06/019. arXiv:2110.08189 [astro-ph.CO]
74. F. Zhang, J. Lin, Y. Lu, Phys. Rev. D **104**, 063515 (2021) [Erratum: Phys.Rev.D 104, 129902 (2021)]. https://doi.org/10.1103/PhysRevD.104.063515. arXiv:2106.10792 [gr-qc]
75. K.T. Abe, R. Inui, Y. Tada, S. Yokoyama, JCAP **05**, 044 (2023). https://doi.org/10.1088/1475-7516/2023/05/044. arXiv:2209.13891 [astro-ph.CO]
76. S. Garcia-Saenz, L. Pinol, S. Renaux-Petel, D. Werth, JCAP **03**, 057 (2023). https://doi.org/10.1088/1475-7516/2023/03/057. arXiv:2207.14267 [astro-ph.CO]
77. J.-P. Li, S. Wang, Z.-C. Zhao, K. Kohri, JCAP **10**, 056 (2023). https://doi.org/10.1088/1475-7516/2023/10/056. arXiv:2305.19950 [astro-ph.CO]
78. G. Franciolini, A. Iovino Jr., V. Vaskonen, H. Veermae, Phys. Rev. Lett. **131**, 201401 (2023). https://doi.org/10.1103/PhysRevLett.131.201401. arXiv:2306.17149 [astro-ph.CO]
79. S. Choudhury, K. Dey, A. Karde, S. Panda, M. Sami, (2023). arXiv:2310.11034 [astro-ph.CO]
80. V. De Luca, G. Franciolini, A. Riotto, Phys. Rev. Lett. **126**, 041303 (2021). https://doi.org/10.1103/PhysRevLett.126.041303. arXiv:2009.08268 [astro-ph.CO]
81. V. Vaskonen, H. Veermäe, Phys. Rev. Lett. **126**, 051303 (2021). https://doi.org/10.1103/PhysRevLett.126.051303. arXiv:2009.07832 [astro-ph.CO]
82. K. Kohri, T. Terada, Phys. Lett. B **813**, 136040 (2021). https://doi.org/10.1016/j.physletb.2020.136040. arXiv:2009.11853 [astro-ph.CO]
83. G. Domènech, S. Pi, Sci. China Phys. Mech. Astron. **65**, 230411 (2022). https://doi.org/10.1007/s11433-021-1839-6. arXiv:2010.03976 [astro-ph.CO]
84. G. Agazie et al. (NANOGrav), Astrophys. J. Lett. **951** (2023). https://doi.org/10.3847/2041-8213/acdac6. arXiv:2306.16213 [astro-ph.HE]
85. G. Agazie et al. (NANOGrav), Astrophys. J. Lett. **951** (2023). https://doi.org/10.3847/2041-8213/acda9a. arXiv:2306.16217 [astro-ph.HE]
86. A. Zic et al., Publ. Astron. Soc. Austral. **40**, e049 (2023). https://doi.org/10.1017/pasa.2023.36. arXiv:2306.16230 [astro-ph.HE]
87. J. Antoniadis et al. (EPTA), Astron. Astrophys. **678**, A48 (2023). https://doi.org/10.1051/0004-6361/202346841. arXiv:2306.16224 [astro-ph.HE]
88. J. Antoniadis et al. (EPTA, InPTA), Astron. Astrophys. **678**, A50 (2023). https://doi.org/10.1051/0004-6361/202346844. arXiv:2306.16214 [astro-ph.HE]
89. H. Xu et al., Res. Astron. Astrophys. **23**, 075024 (2023). https://doi.org/10.1088/1674-4527/acdfa5. arXiv:2306.16216 [astro-ph.HE]
90. Y.-F. Cai, X.-C. He, X. Ma, S.-F. Yan, G.-W. Yuan, (2023). arXiv:2306.17822 [gr-qc]
91. S. Wang, Z.-C. Zhao, J.-P. Li, Q.-H. Zhu, Phys. Rev. Res. **6**, L012060 (2024). https://doi.org/10.1103/PhysRevResearch.6.L012060. arXiv:2307.00572 [astro-ph.CO]
92. Z. Yi, Q. Gao, Y. Gong, Y. Wang, F. Zhang, Sci. China Phys. Mech. Astron. **66**, 120404 (2023). https://doi.org/10.1007/s11433-023-2266-1. arXiv:2307.02467 [gr-qc]
93. Z. Yi, Z.-Q. You, Y. Wu, JCAP **01**, 066 (2024). https://doi.org/10.1088/1475-7516/2024/01/066. arXiv:2308.05632 [astro-ph.CO]
94. K. Harigaya, K. Inomata, T. Terada, Phys. Rev. D **108**, 123538 (2023). https://doi.org/10.1103/PhysRevD.108.123538. arXiv:2309.00228 [astro-ph.CO]
95. L. Liu, Z.-C. Chen, Q.-G. Huang, Phys. Rev. D **109**, L061301 (2024). https://doi.org/10.1103/PhysRevD.109.L061301. arXiv:2307.01102 [astro-ph.CO]
96. Z.-C. Chen, L. Liu, (2024). arXiv:2402.16781 [astro-ph.CO]
97. L. Liu, Y. Wu, Z.-C. Chen, JCAP **04**, 011 (2024). https://doi.org/10.1088/1475-7516/2024/04/011. arXiv:2310.16500 [astro-ph.CO]
98. K. Danzmann, Class. Quantum Gravity **14**, 1399 (1997). https://doi.org/10.1088/0264-9381/14/6/002
99. P. Amaro-Seoane et al. (LISA), (2017). arXiv:1702.00786 [astro-ph.IM]
100. S. Kawamura et al., PTEP **2021**, 05A105 (2021). https://doi.org/10.1093/ptep/ptab019. arXiv:2006.13545 [gr-qc]
101. W.-R. Hu, Y.-L. Wu, Natl. Sci. Rev. **4**, 685 (2017). https://doi.org/10.1093/nsr/nwx116
102. J. Luo et al. (TianQin), Class. Quantum Gravity **33**, 035010 (2016). https://doi.org/10.1088/0264-9381/33/3/035010. arXiv:1512.02076 [astro-ph.IM]
103. Y. Gong, J. Luo, B. Wang, Nat. Astron. **5**, 881 (2021). https://doi.org/10.1038/s41550-021-01480-3. arXiv:2109.07442 [astro-ph.IM]
104. P. Horava, Phys. Rev. D **79**, 084008 (2009). https://doi.org/10.1103/PhysRevD.79.084008. arXiv:0901.3775 [hep-th]
105. M. Crisostomi, K. Noui, C. Charmousis, D. Langlois, Phys. Rev. D **97**, 044034 (2018). https://doi.org/10.1103/PhysRevD.97.044034. arXiv:1710.04531 [hep-th]
106. X. Gao, X.-Y. Hong, Phys. Rev. D **101**, 064057 (2020). https://doi.org/10.1103/PhysRevD.101.064057. arXiv:1906.07131 [gr-qc]
107. Y.-M. Hu, X. Gao, Phys. Rev. D **105**, 044023 (2022). https://doi.org/10.1103/PhysRevD.105.044023. arXiv:2111.08652 [gr-qc]
108. Y.-M. Hu, X. Gao, Phys. Rev. D **104**, 104007 (2021). https://doi.org/10.1103/PhysRevD.104.104007. arXiv:2104.07615 [gr-qc]
109. T. Zhu, W. Zhao, A. Wang, Phys. Rev. D **107**, 044051 (2023). https://doi.org/10.1103/PhysRevD.107.044051. arXiv:2211.04711 [gr-qc]
110. T. Zhu, W. Zhao, A. Wang, Phys. Rev. D **107**, 024031 (2023). https://doi.org/10.1103/PhysRevD.107.024031. arXiv:2210.05259 [gr-qc]
111. S. Akama, M. Zhu, (2024). arXiv:2404.05464 [gr-qc]
112. W. Zhao, T. Liu, L. Wen, T. Zhu, A. Wang, Q. Hu, C. Zhou, Eur. Phys. J. C **80**, 630 (2020). https://doi.org/10.1140/epjc/s10052-020-8211-4. arXiv:1909.13007 [gr-qc]

113. Y.-F. Wang, R. Niu, T. Zhu, W. Zhao, Astrophys. J. **908**, 58 (2021). https://doi.org/10.3847/1538-4357/abd7a6. arXiv:2002.05668 [gr-qc]
114. C. Gong, T. Zhu, R. Niu, Q. Wu, J.-L. Cui, X. Zhang, W. Zhao, A. Wang, Phys. Rev. D **105**, 044034 (2022). https://doi.org/10.1103/PhysRevD.105.044034. arXiv:2112.06446 [gr-qc]
115. T.-C. Li, T. Zhu, W. Zhao, A. Wang, (2024). arXiv:2403.05841 [gr-qc]
116. R. Jackiw, S.Y. Pi, Phys. Rev. D **68**, 104012 (2003). https://doi.org/10.1103/PhysRevD.68.104012. arXiv:gr-qc/0308071
117. M. Zhu, Y. Cai, JHEP **04**, 095 (2023). https://doi.org/10.1007/JHEP04(2023)095. arXiv:2301.13502 [gr-qc]
118. Z. Li, J. Qiao, T. Liu, T. Zhu, W. Zhao, JCAP **04**, 006 (2023). https://doi.org/10.1088/1475-7516/2023/04/006. arXiv:2211.12188 [gr-qc]
119. Z. Li, J. Qiao, T. Liu, S. Hou, T. Zhu, W. Zhao, (2023). arXiv:2309.05991 [gr-qc]
120. C. Fu, J. Liu, T. Zhu, H. Yu, P. Wu, Eur. Phys. J. C **81**, 204 (2021). https://doi.org/10.1140/epjc/s10052-021-09001-2. arXiv:2006.03771 [gr-qc]
121. S.D. Odintsov, V.K. Oikonomou, F.P. Fronimos, Phys. Dark Univ. **35**, 100950 (2022). https://doi.org/10.1016/j.dark.2022.100950. arXiv:2108.11231 [gr-qc]
122. S. Nojiri, S.D. Odintsov, V.K. Oikonomou, A.A. Popov, Phys. Rev. D **100**, 084009 (2019). https://doi.org/10.1103/PhysRevD.100.084009. arXiv:1909.01324 [gr-qc]
123. S. Nojiri, S.D. Odintsov, V.K. Oikonomou, A.A. Popov, Phys. Dark Univ. **28**, 100514 (2020). https://doi.org/10.1016/j.dark.2020.100514. arXiv:2002.10402 [gr-qc]
124. S.D. Odintsov, V.K. Oikonomou, EPL **129**, 40001 (2020). https://doi.org/10.1209/0295-5075/129/40001. arXiv:2003.06671 [gr-qc]
125. S.D. Odintsov, V.K. Oikonomou, R. Myrzakulov, Symmetry **14**, 729 (2022). https://doi.org/10.3390/sym14040729. arXiv:2204.00876 [gr-qc]
126. A. Nishizawa, T. Kobayashi, Phys. Rev. D **98**, 124018 (2018). https://doi.org/10.1103/PhysRevD.98.124018. arXiv:1809.00815 [gr-qc]
127. J. Qiao, T. Zhu, W. Zhao, A. Wang, Phys. Rev. D **101**, 043528 (2020). https://doi.org/10.1103/PhysRevD.101.043528. arXiv:1911.01580 [astro-ph.CO]
128. J. Qiao, T. Zhu, W. Zhao, A. Wang, Phys. Rev. D **100**, 124058 (2019). https://doi.org/10.1103/PhysRevD.100.124058. arXiv:1909.03815 [gr-qc]
129. J. Qiao, T. Zhu, G. Li, W. Zhao, JCAP **04**, 054 (2022). https://doi.org/10.1088/1475-7516/2022/04/054. arXiv:2110.09033 [gr-qc]
130. J. Qiao, Z. Li, T. Zhu, R. Ji, G. Li, W. Zhao, Front. Astron. Space Sci. **9**, 1109086 (2023). https://doi.org/10.3389/fspas.2022.1109086. arXiv:2211.16825 [gr-qc]
131. N. Bartolo, L. Caloni, G. Orlando, A. Ricciardone, JCAP **03**, 073 (2021). https://doi.org/10.1088/1475-7516/2021/03/073. arXiv:2008.01715 [astro-ph.CO]
132. T. Zhu, W. Zhao, J.-M. Yan, C. Gong, A. Wang, (2023). arXiv:2304.09025 [gr-qc]
133. J. Qiao, Z. Li, R. Ji, T. Zhu, G. Li, W. Zhao, J. Chen, JCAP **10**, 066 (2023). https://doi.org/10.1088/1475-7516/2023/10/066. arXiv:2307.12886 [gr-qc]
134. F. Zhang, J.-X. Feng, X. Gao, JCAP **10**, 054 (2022). https://doi.org/10.1088/1475-7516/2022/10/054. arXiv:2205.12045 [gr-qc]
135. J.-X. Feng, F. Zhang, X. Gao, JCAP **07**, 047 (2023). https://doi.org/10.1088/1475-7516/2023/07/047. arXiv:2302.00950 [gr-qc]
136. F. Zhang, J.-X. Feng, X. Gao, Phys. Rev. D **108**, 063513 (2023). https://doi.org/10.1103/PhysRevD.108.063513. arXiv:2307.00330 [gr-qc]
137. F. Zhang, J.-X. Feng, X. Gao, (2024). arXiv:2404.02922 [gr-qc]
138. C. Tzerefos, T. Papanikolaou, E.N. Saridakis, S. Basilakos, Phys. Rev. D **107**, 124019 (2023). https://doi.org/10.1103/PhysRevD.107.124019. arXiv:2303.16695 [gr-qc]
139. S. Garcia-Saenz, Y. Lu, Z. Shuai, Phys. Rev. D **108**, 123507 (2023). https://doi.org/10.1103/PhysRevD.108.123507. arXiv:2306.09052 [gr-qc]
140. S. Alexander, N. Yunes, Phys. Rept. **480**, 1 (2009). https://doi.org/10.1016/j.physrep.2009.07.002. arXiv:0907.2562 [hep-th]
141. N. Bartolo, G. Orlando, JCAP **1707**, 034 (2017). https://doi.org/10.1088/1475-7516/2017/07/034. arXiv:1706.04627 [astro-ph.CO]
142. N. Bartolo, G. Orlando, M. Shiraishi, JCAP **1901**, 050 (2019). https://doi.org/10.1088/1475-7516/2019/01/050. arXiv:1809.11170 [astro-ph.CO]
143. B.P. Abbott et al. (LIGO Scientific, Virgo, Fermi-GBM, INTEGRAL), Astrophys. J. Lett. **848**, L13 (2017). https://doi.org/10.3847/2041-8213/aa920c. arXiv:1710.05834 [astro-ph.HE]
144. S. Saito, K. Ichiki, A. Taruya, JCAP **09**, 002 (2007). https://doi.org/10.1088/1475-7516/2007/09/002. arXiv:0705.3701 [astro-ph]
145. V. Gluscevic, M. Kamionkowski, Phys. Rev. D **81**, 123529 (2010). https://doi.org/10.1103/PhysRevD.81.123529. arXiv:1002.1308 [astro-ph.CO]
146. K. Inomata, M. Kawasaki, K. Mukaida, Y. Tada, T.T. Yanagida, Phys. Rev. D **95**, 123510 (2017). https://doi.org/10.1103/PhysRevD.95.123510. arXiv:1611.06130 [astro-ph.CO]